\documentclass[final,3p,times,twocolumn]{elsarticle}
\usepackage[utf8]{inputenc}
\usepackage[dvipsnames]{xcolor}
\usepackage{xspace}
\usepackage{makecell}
\usepackage{rotating}
\usepackage{ulem}
\usepackage{amssymb}
\usepackage{amsmath}
\newcommand{\gx}{GX\,301$-$2\xspace}
\newcommand{\sd}{$'$\,}
\newcommand{\dd}{$''$\,}

\newcommand{\markp}[1]{\textcolor{black} { #1}}
\usepackage{lineno}

\journal{Astroparticle Physics}

\begin{document}
\begin{frontmatter}


%
%
\title{{\bf {\textit {XL-Calibur}} -- a second-generation balloon-borne hard X-ray polarimetry mission}}
%
%
\author[WUSTL]{Q.\,Abarr}
\author[Ehime]{H.\,Awaki}
\author[RICE]{M.G.\,Baring}
\author[WUSTL]{R.\,Bose}
\author[DGC]{G.\,De Geronimo}
\author[WUSTL]{P.\,Dowkontt}
\author[WUSTL]{M.\,Errando}
\author[GE]{V.\,Guarino}
\author[Osaka]{K.\,Hattori}
\author[Osaka,ISAS]{K.\,Hayashida}
\author[Hiroshima]{F.\,Imazato}
\author[ISAS]{M.\,Ishida}
\author[KTH1,KTH2]{N.K.\,Iyer\corref{cor1}}
\author[UNH]{F.\,Kislat\corref{cor1}}
\author[KTH1,KTH2]{M.\,Kiss}
\author[RIKEN1,RIKEN2]{T.\,Kitaguchi}
\author[WUSTL]{H.\,Krawczynski\corref{cor1}}
\author[WUSTL]{L.\,Lisalda\corref{cor1}} 
\author[Hiroshima]{H.\,Matake}
\author[ISAS]{Y.\,Maeda}
\author[Osaka]{H.\,Matsumoto}
\author[Osaka]{T.\,Mineta}
\author[OIST]{T.\,Miyazawa}
\author[Hiroshima]{T.\,Mizuno}
\author[GSFC]{T.\,Okajima}
\author[KTH1,KTH2]{M.\,Pearce\corref{cor1}}
\author[WUSTL]{B.F.\,Rauch}
\author[KTH1,KTH2]{F.\,Ryde}
\author[WFF]{C.\,Shreves}
\author[UNH]{S.\,Spooner}
\author[KTH1,KTH2]{T.-A.\,Stana}
\author[Hiroshima]{H.\,Takahashi}
\author[TokyoMet]{M.\,Takeo}
\author[RIKEN1,RIKEN2]{T.\,Tamagawa}
\author[GSFC]{K.\,Tamura}
\author[Osaka]{H.\,Tsunemi}
\author[Hiroshima]{N.\,Uchida}
\author[Hiroshima]{Y.\,Uchida}
\author[WUSTL]{A.T.\,West}
\author[NRL]{E.A.\,Wulf}
\author[Hiroshima]{R.\,Yamamoto}
\cortext[cor1]{Corresponding authors: N.K.\,Iyer ({\it nkiyer@kth.se}), F.\,Kislat ({\it fabian.kislat@unh.edu}), H.\,Krawczynski ({\it krawcz@wustl.edu}), L.\,Lisalda ({\it lindsey.lisalda@wustl.edu}), M.\,Pearce ({\it pearce@kth.se}).}
\address[WUSTL]{Washington University in St. Louis, 1 Brookings Drive, CB 1105, St. Louis, MO 63130, USA.}
\address[Ehime]{Graduate School of Science and Engineering, Ehime University, Bunkyo-cho, Matsuyama, Ehime, Japan.}
\address[RICE]{Rice University, Department of Physics and Astronomy, 6100 Main Street, Houston, TX 77251, USA.}
\address[DGC]{DG CIRCUITS, 30 Pine Road, Syosset, NY 11791, USA.}
\address[GE]{Guarino Engineering Services, 1134 S Scoville Avenue, Oak Park, IL 60304, USA.}
\address[Osaka]{Osaka University, Department of Earth and Space Science, Graduate School of Science,
and Project Research Center for Fundamental Sciences, 1-1 Machikaneyama-cho, Toyonaka, Osaka 560-0043, Japan.}
\address[ISAS]{ISAS, 3-1-1 Yoshinodai, Sagamihara, Kanagawa 229-8510, Japan.}
\address[Hiroshima]{Hiroshima University, 1-3-1 Kagamiyama, Higashi-Hiroshima, Hiroshima 739-8526, Japan.}
\address[KTH1]{KTH Royal Institute of Technology, Department of Physics, 106 91 Stockholm, Sweden.}
\address[KTH2]{The Oskar Klein Centre for Cosmoparticle Physics, AlbaNova University Centre, 106 91 Stockholm, Sweden.}
\address[UNH]{University of New Hampshire, Department of Physics and Astronomy, and Space Science Center, Morse Hall, 8 College Road, Durham, NH 03824, USA.}
\address[RIKEN1]{RIKEN Cluster for Pioneering Research, 2-1 Hirosawa, Wako, Saitama 351-0198, Japan.}
\address[RIKEN2]{RIKEN Nishina Center, 2-1 Hirosawa, Wako, Saitama 351-0198, Japan.}
\address[OIST]{Okinawa Institute of Science and Technology Graduate University, Kunigami-gun, Japan.}
\address[GSFC]{NASA Goddard Space Flight Center, Greenbelt, MD 20771, USA.}
\address[WFF]{NASA Wallops Flight Facility, 32400 Fulton Street, Wallops Island, VA 23337, USA.}
\address[TokyoMet]{Tokyo Metropolitan University, 1-1 Minami-Osawa, Hachioji, Tokyo 192-0397, Japan.}
\address[NRL]{U.S. Naval Research Laboratory, 4555 Overlook Avenue, SW Washington, DC 20375, USA.}

%
%
\begin{abstract}
{\it XL-Calibur} is a hard X-ray (15-80~keV) polarimetry mission operating from a stabilised balloon-borne platform in the stratosphere. 
It builds on heritage from the {\it X-Calibur} mission, which observed the accreting neutron star \gx  from Antarctica, between December 29th 2018 and January 1st 2019.
The {\it XL-Calibur} design incorporates an X-ray mirror, which focusses X-rays onto a polarimeter comprising a beryllium rod surrounded by Cadmium Zinc Telluride (CZT) detectors. 
The polarimeter is housed in an anticoincidence shield to mitigate background from particles present in the stratosphere.
The mirror and polarimeter-shield assembly are mounted at opposite ends of a 12~m long lightweight truss, which is pointed with arcsecond precision by WASP -- the Wallops Arc Second Pointer.   
The {\it XL-Calibur} mission will achieve a substantially improved sensitivity over {\it X-Calibur} by using a larger effective area X-ray mirror, reducing background through thinner CZT detectors, and improved anticoincidence shielding.
When observing a 1 Crab source for $t_{\rm day}$ days, the Minimum Detectable Polarisation (at 99\% confidence level) 
is $\sim$2$\%\cdot t_{\rm day}^{-1/2}$. 
\markp{The energy resolution at 40~keV is $\sim$5.9~keV.}
The aim of this paper is to describe the design and performance of the {\it XL-Calibur} mission, as well as the foreseen science programme. 
\end{abstract}
%
%
\begin{keyword}
X-ray polarimetry \sep scientific ballooning \sep compact objects 
\end{keyword}
\end{frontmatter}
%
%
\section{Introduction}
Black-hole systems, neutron stars and other compact objects are too small and distant to be imaged. Information on source geometry and high-energy emission mechanisms is instead derived from spectral and timing measurements. Although spectacular advances have been made, results are often model-dependent with interpretation subject to irresolvable degeneracies. X-ray polarimetry provides an independent diagnostic, which probes anisotropies due to relativistic motions and/or the presence of magnetic fields in sources. Two new observables are introduced to describe the high-energy emission -- the linear polarisation fraction (\%) and the linear polarisation angle (degrees). One of the highlights of astrophysics during this decade will be establishing X-ray polarimetry as a new window on the high-energy universe. 

\begin{figure*}[t!]
\begin{center}
    \includegraphics[width=0.60\linewidth]{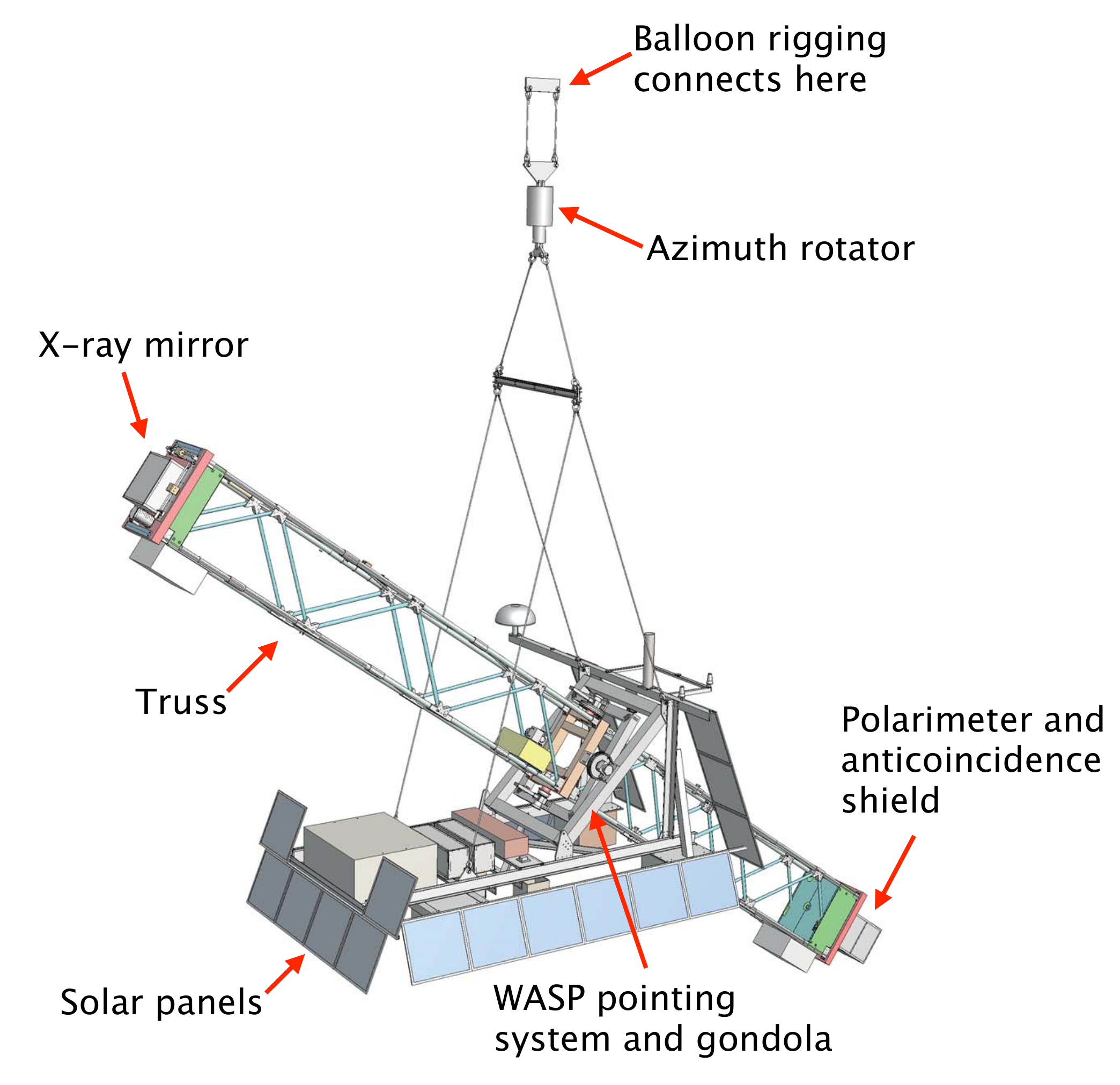}
\end{center}
\caption{\label{f:xlc} The {\it XL-Calibur} mission will use a 12~m long truss equipped with an X-ray mirror at one end (image left) and a scattering polarimeter at the other end (image right). The {\it XL-Calibur} truss will use the same design elements as the 8 m long {\it X-Calibur} truss, but with larger-diameter and thicker-wall carbon fibre tubes and Al joints with increased strength to obtain a similar overall stiffness. The resulting X-ray telescope is pointed
by the WASP system with an absolute pointing knowledge of 15\dd and a pointing precision of $<$1\dd Root Mean Square (RMS).
}
\end{figure*}

In the soft X-ray band (2--8 keV), a major step forward will be provided by the Imaging X-ray Polarimetry Explorer (IXPE) satellite mission which is scheduled for launch in 2021~\cite{2016SPIE.9905E..17W,2018Galax...6...33W}. 
In the hard X-ray band, observations are possible from the stratosphere ($\sim$40~km altitude) and balloon-borne polarimeters have recently made initial observations of bright sources in the $\sim$(15--100~keV) energy band~\cite{Abarr_2020,2017NatSR...7.7816C,2018NatAs...2..652C}.
This paper describes the design of a second-generation balloon-borne mission, {\it XL-Calibur} (XL stands for eXtra Large), shown in Fig.~\ref{f:xlc}, which will greatly extend polarimetric measurements in the 15-80~keV band. 
{\it XL-Calibur}  follows on from the {\it X-Calibur} mission~\cite{2011APh....34..550K,2013APh....41...63G,2014JAI.....340008B,2017JAI.....640003K,2018JATIS...4a1004K}, which was flown on two engineering flights from Fort Sumner, New Mexico, USA (2014 \& 2016) and as a Long Duration Balloon flight from McMurdo, Antarctica (December 2018--January 2019). 
Although the Antarctica flight was unexpectedly brief ($\sim$2 days long), {\it X-Calibur} made detailed temporal and spectral observations of the accretion-powered pulsar \gx, and constrained polarisation parameters~\cite{Abarr_2020}. 
The {\it X-Calibur} observations were complemented by simultaneous spectral and timing studies by {\it NICER}, {\it Swift} XRT, and {\it Fermi} GBM.

The {\it XL-Calibur} mission uses a 12~m focal length X-ray mirror to focus X-rays onto an actively shielded polarimeter comprising a beryllium scattering rod surrounded by Cadmium Zinc Telluride (CZT) detectors. The mirror and polarimeter assemblies are mounted at either end of a lightweight truss, which can be pointed with arcsecond precision.
{\it XL-Calibur} will replace the {\it InFOC$\mu$S} 8~m focal length mirror~\cite{IF1,IF2,IF3} 
used by {\it X-Calibur} with the 12~m focal length mirror from the Formation Flight Astronomical Survey Telescope ({\it FFAST}) mission~\cite{2014SPIE.9144E..2RT}, thereby achieving a 3 (10) times larger collection area at 15\,keV (60\,keV). 
Compared to {\it X-Calibur}, {\it XL-Calibur} will benefit from a background count rate reduced by a factor of 25 through a combination of thinner CZT detectors and improved anticoincidence shielding. 
The {\it XL-Calibur} technique is readily transferrable to a satellite platform~\cite{2019arXiv190710190J,2016APh....75....8K}.
Table~\ref{table:leads} details the {\it XL-Calibur} team leads.

Combining observations of future soft X-ray polarimeters like {\it IXPE}, {\it eXTP}~\cite{2019SCPMA..6229502Z}, 
{\it PRAXyS}~\cite{2016NIMPA.838...89I}, or {\it REDSOX}~\cite{2018JATIS...4a1005M} with those of the hard X-ray polarimeter {\it XL-Calibur} will be a cost-effective option for harvesting some of the science highlights of X-ray polarimetry. 
There is also synergy with proposed wide field-of-view hard X-ray polarimetry missions such as COSI~\cite{COSI}, LEAP~\cite{LEAP} and POLAR--2~\cite{POLAR2}.
The polarimetric observations will provide geometric information on emission regions a few femto-degrees across (for a source at the distance of the black-hole binary Cyg\,X$-$1). 
Joint measurements of the temporal, spectral and linear polarisation properties of the emission from neutron stars and black-hole systems will probe strong gravity, strong-field quantum electrodynamics (QED), and the behaviour of hadronic matter at extreme densities and pressures. 
While broadband observations are important for X-ray timing and spectral studies, 
they are essential for polarisation studies where the change of polarisation fraction and angle with energy, rather than the absolute values at specific energies, reveals the geometry and physical properties of the emission region.

%
%

Two {\it XL-Calibur} flights have been approved under the NASA Astrophysics Research and Analysis (APRA) programme. 
The first flight will take place from Esrange, Sweden (to Canada, 5-7 day flight) in mid-2022. 
The second flight is foreseen from McMurdo, Antarctica (circumpolar, $\sim$8-55 day flight), nominally at the end of 2023. 
The 15-80 keV {\it XL-Calibur} observations will be highly complementary to the 2--8 keV {\it IXPE} observations and allow the energy dependence of polarisation parameters to be studied. 
An overview of the science drivers for X-ray polarimetry is provided in
\cite{2011APh....34..550K,1997SSRv...82..309L,2006astro.ph.11483W,2010xpnw.book.....B,2013ExA....36..523S,2019arXiv190409313K}.
The highlights of the {\it XL-Calibur} science programme are as follows:  
\begin{enumerate}
\item
{\it XL-Calibur} observations of the hard X-ray emission of stellar-mass black holes in X-ray binaries such as Cyg\,X$-$1 and GX\,339$-$4 will constrain the properties of the X-ray bright coronas. The joint {\it IXPE} and {\it XL-Calibur} results will disentangle the polarisation of the thermal accretion disk emission, and the direct and reflected coronal emission. 
\item {\it XL-Calibur} is ideally suited to make precision measurements of the birefringent properties of the QED vacuum surrounding highly-magnetised accreting pulsars like Her\,X$-$1, \gx, and Vela\,X$-$1. This is particularly informative at the energies of their
Cyclotron Resonant Scattering Features (CRSF), where the competition of QED and plasma birefringence is expected to lead to extremely high and strongly energy-dependent polarisation. 
The joint {\it IXPE} and {\it XL-Calibur} observations have the potential to determine the emission
geometry (e.g.\ pencil beam or fan beam) and to 
study the impact of strong QED effects on the birefringence of the magnetised plasma and the polarisation-dependent scattering cross-sections.
\item {\it XL-Calibur} can distinguish between competing emission models of the rotation-powered Crab pulsar -- an archetypical cosmic particle accelerator. {\it XL-Calibur}'s high sensitivity will allow phase-resolved polarimetry, cleanly separating contributions from the pulsar and from the nebula.
\end{enumerate}
\begin{table*}[tbh]
\begin{center}
\begin{tabular}{|l|l|p{8cm}|}
\hline
\textbf{Name} & \textbf{Affiliation} & \textbf{Role}\\
\hline
H. Krawczynski$^*$& WUSTL & Principal Investigator, science analysis, polarimeter, truss fabrication and test \\
H. Awaki 		& Ehime University & Mirror mounting on gondola \\
R. Bose 		& WUSTL & Electronics lead \\
D. Braun 		& WUSTL & Component design \\
G. De Geronimo 	& DG CIRCUITS & Polarimeter ASICs \\
V. Guarino 		& Guarino Engineering & Mechanical design, including truss \\ 
K. Harmon 		& NASA WFF & WASP management \\ 
S. Heatwole 		& NASA WFF & Star cameras \\
M. Ishida 		& ISAS/JAXA & Mirror alignment bars \\
F. Kislat$^*$ 	& University of New Hampshire & Data acquisition and telemetry software, science simulations and analysis \\
J. Lanzi 		& NASA WFF & WASP design \\
Y. Maeda$^*$ 	& ISAS/JAXA & Mirror fabrication, calibration and alignment \\   
H. Matsumoto 	& Osaka University & Mirror calibration with SPring-8 synchrotron beams \\
T. Okajima$^*$ 	& NASA GSFC & Mirror alignment \\
M. Pearce$^*$ 	& KTH & Lead of Swedish team -- background simulations, BGO anticoincidence shield, science analysis\\
H. Takahashi$^*$ 	& Hiroshima University & Lead of Japanese team. Support to mirror activities, science analysis \\  
E.A. Wulf 		& NRL & Polarimeter ASICs \\
\hline
\end{tabular}
\end{center}
\caption{\label{table:leads} {\it XL-Calibur} team leads. The persons marked with an asterisk form the {\it XL-Calibur} Executive Committee, which assists the Principal Investigator with mission management.}
\end{table*}

\begin{figure*}[t]
\begin{center}
\includegraphics[width=.99\linewidth]{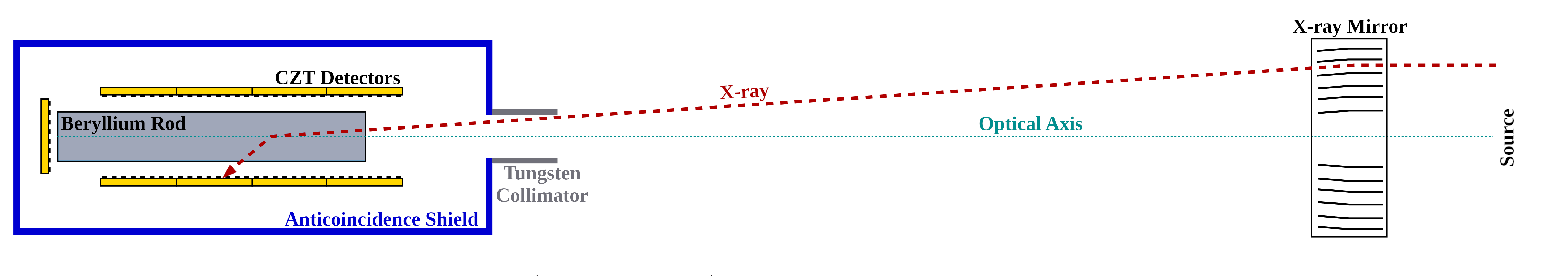}
\end{center}
\caption{\label{f:dp} The {\it XL-Calibur} detection principle (not drawn to scale): the X-ray mirror focusses source photons onto a 
beryllium scattering rod located 12~m away. A scattered photon is detected in the surrounding assembly of CZT detectors.
The distribution of the azimuthal scattering angles is used to measure the linear polarisation fraction and angle. 
A CZT detector is mounted at the far end of the beryllium rod to allow alignment studies during flight (see Section~\ref{ss:poldesign}).}
\end{figure*}

The remainder of the paper is structured as follows. 
The design of the {\it XL-Calibur} mission is detailed in Section~\ref{s:design}, the expected performance is presented in Section~\ref{sec:performance}, the science programme is described in Section~\ref{s:science}, and a discussion is presented in Section~\ref{s:discussion}.

%
%
\section{Mission Design and Implementation}
\label{s:design}
{\it XL-Calibur} uses a 12~m focal length mirror to focus X-rays onto an actively shielded broadband scattering polarimeter. The components are mounted on a truss, which is pointed with arcsecond precision using the Wallops Arc Second Pointer (WASP)~\cite{Stuc:15}. Focussed X-rays impinge on the centre of a beryllium rod (Fig.~\ref{f:dp}). 
Owing to the low atomic number of beryllium, a large fraction (e.g. {$\sim$85\%} at 30~keV) of the X-rays scatter from the rod into a circumjacent assembly of high atomic number CZT detectors. 
As linearly polarised X-rays scatter preferentially perpendicular to the orientation of the electric field vector, the distribution of azimuthal scattering angles encodes the polarisation fraction and angle. 
For a beam with polarisation fraction, $p_0$, and polarisation angle, $\psi_0$, 

\begin{equation}
\frac{dN}{d\psi} \,=\, \frac{1}{2\pi} \left[1+ \mu\, p_0 \,\cos{(2 (\psi-\psi_0 - \pi/2))}\right],
\end{equation}
where $N$ is the number of photons scattered, $\psi$ is the azimuthal scattering angle, and $\mu\,=\,51.3\%$ is the modulation factor evaluated for {\it X-Calibur} (also representative for {\it XL-Calibur}). 
The modulation factor is largely energy-independent across the {\it XL-Calibur} energy range~\cite{2014JAI.....340008B}. 
The mirror focusses X-rays using grazing incidence reflection, which reduces the polarisation fraction by less than 1\%~\cite{1993ApOpt..32.4231S,2009NIMPA.603..393K}.
Table \ref{t:xlc} summarises the {\it XL-Calibur} design and performance parameters, including the minimum detectable polarisation (MDP, \%)~\cite{2006astro.ph.11483W}, 
where there is a 1\% chance to measure a polarisation fraction $\ge\!\mathrm{MDP}$ for an unpolarised beam, and, 

\begin{equation}
    {\rm MDP}\,=\,\frac{429\%}{\mu R_S}
    \sqrt{\frac{R_S+R_{BG}}{t_{\rm obs}}},
\end{equation}
$t_{\rm obs}$ is the on-source integration time in seconds (expressed in days, $t_{\rm day}$, in Table~\ref{t:xlc}), and $R_S$ ($R_{BG}$) is the source (background) counting rate (Hz).

\begin{sidewaystable*}[!htbp]
    \begin{tabular}{|p{2.6cm}|p{6.1cm}|p{8.1cm}|}
    \hline
      \textbf{Component} & \textbf{Description} & \textbf{Parameters}\\
      \hline
Truss  & Carbon fibre tubes and aluminium joints & Focal spot movement: $<$3 mm \\
Pointing system & Pitch-yaw articulated & Pointing precision: 1\dd RMS \\
Star camera & 100~mm, f/1.5 short-wave infrared lens & Pointing knowledge: $<$15\dd (3$\sigma$)\\
X-ray mirror & Wolter I, 12 m focal length, diameter 45~cm, 213 Pt-C coated shells & Effective area: 180 cm$^2$ at 30 keV \\
Polarimeter & Beryllium scatterer, 17~CZT detectors (each: 0.8$\times$20$\times$20 mm$^3$, 64 pixels), NRL1 ASIC readout & Bandpass: 15-80 keV; \markp{$\Delta E$(40 keV)=5.9~keV FWHM}  \\ 
\hline
Power & Science payload and WASP & 460~W\\
Mass & Total mass suspended under rotator & 2132 kg (4700 lbs) \\ 
\hline
\hline
\multicolumn{3}{|c|}{Performance assuming: altitude, 38.1 km (125\,kft); energy range, 15-80~keV}  \\
\hline
Signal rate & 1 Crab source at 60$^{\circ}$ elevation  & 3.3 Hz \\
\hline
Background rate & 100~keV shield veto threshold & 
\begin{tabular}{c|c|c}
{\bf X-Calibur shield} & {\bf New BGO shield} & \\
\hline
2.9~Hz & 0.6~Hz & Solar min. \\
\cline{1-3}
2.0~Hz & 0.4~Hz & Solar max. \\
\end{tabular}
\\
\cline{1-3}
MDP (99\% CL) & 1 Crab source at 60$^{\circ}$ elevation; Modulation Factor: 0.51& 
\begin{tabular}{c|c|c}
{\bf X-Calibur shield} & {\bf New BGO shield} & \\
\hline
{2.1\% $t_{\rm day}^{-1/2}$} & {1.7\% $t_{\rm day}^{-1/2}$} & Solar min. \\
\cline{1-3}
{1.9\% $t_{\rm day}^{-1/2}$} & {1.6\% $t_{\rm day}^{-1/2}$} & Solar max. \\
\end{tabular}
\\
\hline
    \end{tabular}
        \caption{\label{t:xlc} {\it XL-Calibur} specifications and estimated performance.}
\end{sidewaystable*}
%
%
\subsection{The WASP gondola and pointing system} 
The truss assembly is mounted in a custom gondola, which incorporates the WASP pointing system~\cite{Stuc:15}. 
The gondola is suspended beneath a modified NASA rotator, which provides large-angle azimuth targeting and coarse azimuth stabilisation. 
The WASP system points the truss using a pitch/yaw articulated gimbal mounted on the gondola. 
Sub-arcsecond pointing is enabled by the mechanical design of the gimbal hubs, where high-precision angular contact bearings float the rotor-side and stator-side of the hub on a central shaft. 
Small-diameter motors act on the central shafts of each hub through gearboxes to eliminate static friction. 
The shafts in each hub pair are counter-rotated, minimising the residual kinetic friction.
Large-diameter brushless direct-current torque motors act on each control axis. 
The pointing attitude is computed by integrating the attitude angles provided by a gyro-based inertial navigation system (Northrop Grumman LN251). 
Absolute pointing information is derived from a custom star camera.
Control torques are computed using a modified proportional-integral-derivative control law for each axis. 
The quaternion output of the star camera is combined with the integrated attitude solution from the LN251 unit using a 6-state extended Kalman filter. The WASP system pointed {\it X-Calibur} with a Root Mean Square (RMS) precision of $\sim$1\dd during the 2016 and 2018/19 balloon flights~\cite{Abarr:19b}.

The {\it XL-Calibur} WASP configuration will include several upgrades.  
{\it XL-Calibur} will use two star cameras. 
One star camera will be co-aligned with the X-ray mirror (as for {\it X-Calibur}), and the other one will be oriented 25$^{\circ}$ from the pointing axis. 
The second star camera will enable pointing at elevations exceeding 65$^{\circ}$ (through the balloon), and pointing at targets in the presence of stratospheric clouds along the line-of-sight.
A sun sensor with a field-of-view of $40^{\circ}$ will provide absolute pointing information 
of targets close to the Sun (e.g.\ when observing the Crab during a flight from Esrange). 
The WASP simulation model predicts that the 12~m long {\it XL-Calibur} telescope truss can be pointed with a performance comparable to or better than that achieved for the 8\,m long version used by {\it X-Calibur}. 
%
\subsection{The 12~m long telescope truss} 
\begin{figure*}[tb]
\begin{center}
    \includegraphics[width=.90\linewidth]{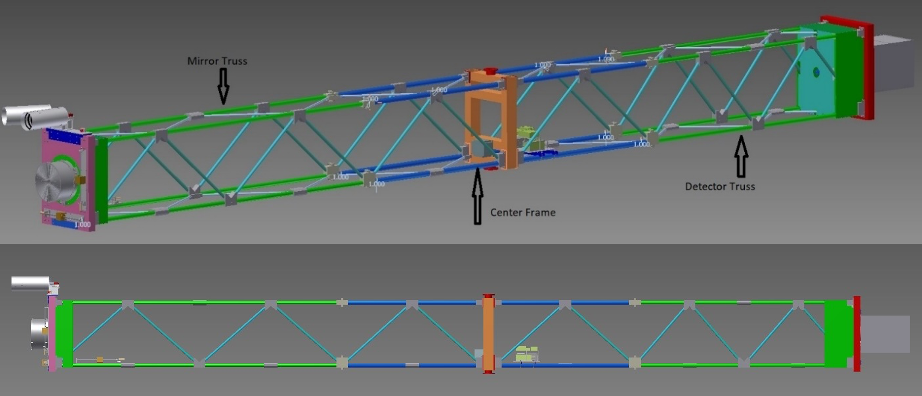}
\end{center}
\caption{\label{f:truss_sections} The design of the 12~m long truss showing the three segments. The polarimeter is located inside the box visible at image-right.}
\end{figure*}

The {\it XL-Calibur} truss will build on the flight-proven design of the {\it X-Calibur} truss~\cite{2017JAI.....640003K}. 
The truss (Fig.~\ref{f:truss_sections}) is composed of five parts, which bolt together: the centre frame of welded aluminium with protruding hubs, which attach to the WASP gondola; the two-part mirror truss, with an aluminium-composite honeycomb panel carrying the 12~m focal length X-ray mirror; and the two-part detector truss, which holds another honeycomb panel, housing the polarimeter-anticoincidence shield assembly. 

Each section comprises carbon fibre tubes with excellent mechanical and thermal expansion properties, which are glued into custom-machined aluminium joints using the epoxy adhesive Loctite E120-HP. A glass-bead bond line controller is added to ensure a uniform bond thickness of 0.018~cm (0.007~inch).

\begin{figure}[tb]
\begin{center}
    \includegraphics[width=.90\linewidth]{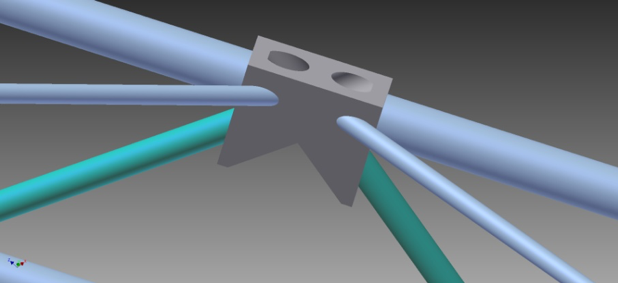}
\end{center}
\caption{\label{f:joint} The {\it XL-Calibur} truss uses carbon fibre tubes joined by aluminium joints. The image shows an example joint. The glue and glass beads mixture is injected through small-diameter holes in the joint.}
\end{figure}

The four main chords of each truss section are continuous throughout the length of the section. Continuous chords allow for fewer joints, increasing the stiffness and reducing thermal deformation of the truss. The aluminium joints at the end of the truss must resist the chord forces while the aluminium joints in the middle of the truss resist shear forces. The detector and mirror trusses use carbon fibre tubes with an outer diameter, OD, of 5.08~cm (2.0~inch). The chords for the centre truss have a larger diameter of 6.35~cm (2.5~inch) OD to reduce deformation and allow the glue joints to resist larger forces. The carbon fibre tubes for the main chords of all three sections have a wall thickness of 0.635~cm (1/4~inch).
The main (side) diagonals are 3.81~cm (1.5~inch) OD tubes with 0.635~cm (1/4~inch) thick walls. The top/bottom lateral diagonals are 2.54~cm (1~inch) OD carbon fibre tubes with 0.32~cm (1/8~inch) thick walls. Smaller-diameter tubes are used in order to minimise the size of the joints, and the thick wall allows equipment to be attached to the truss using U-bolts. The specifications of the carbon fibre tubes are summarised in Table~\ref{t:tubes}, and a typical joint is shown in Fig.~\ref{f:joint}.  

The truss structure must not fail during the large forces experienced during parachute deployment at the end of the flight.
Safety requirements are stipulated by the NASA ballooning 
office\footnote{NASA document {\it Structural Requirements and Recommendations for Balloon Gondola Design}. \\
{\tt https://www.csbf.nasa.gov/docs.html}}.
The truss must be designed to tolerate a 16~$g$ acceleration aligned with the Earth's gravity vector. A large sample of carbon fibre tube-aluminium joints are currently being tested in order to optimise the strength of the joints and to accumulate statistics about sample-to-sample variations. 
Joints where carbon fibre tubes are inserted into aluminium joints are shown to have far superior strength than aluminium lugs, which insert into the bore of carbon fibre tubes, especially after temperature cycling the joints between $-$60\,$^{\circ}$C and +50\,$^{\circ}$C. The former design leads to a compression, and strengthening, of the glue lines when the samples are cooled to the temperatures expected during the ascent in the atmosphere (particularly in the tropopause) and after landing in Antarctica. The latter design leads to an expansion of the glue lines in cold conditions, which would instead weaken the joints. 

For a nominal (1~$g$) load, the maximum tensile force of 678~kg (1494~lbs) is found in the top main chord at the centre support corresponding to a stress of 5.8~MPa (845~psi) in the tube and a glue shear stress of 0.82~MPa (119~psi) -- a factor of 10.6 less than the strength of the main chord. The maximum force in the diagonal is 215~kg (475~lbs), which is a stress of 2.4~MPa (346~psi) in the tube and a shear stress in the adhesive of 0.7~MPa (101~psi) -- a factor of 39 below the strength of the diagonal.  
\begin{table}[t]
  \begin{center}
    \label{t:cf}
    \begin{tabular}{|c|c|}
    \hline
      \textbf{Property} & \textbf{Specification}\\
      \hline
       Density & 1.66 g cm$^{-3}$ \\
       Axial modulus of elasticity & 97 GPa \\
       Ultimate tensile strength & 0.96 GPa \\
       Thermal expansion & 0.18 $\times 10^{-6}$ K$^{-1}$\\ 
      \hline
    \end{tabular}
  \end{center}
        \caption{\label{t:tubes} Specifications of the 
        carbon fiber tubes of the 12~m {\it XL-Calibur} 
        telescope truss.}
\end{table}

The truss will be certified using a combination of methods: $(a)$ systematic tests of joint coupons until failure, prior to truss fabrication; $(b)$ witness samples, produced alongside the flight components and tested until failure; $(c)$ a load test of the assembled truss.
The {\it X-Calibur} truss met the requirement of $<$3 mm focal spot movements during the 2018/2019 Antarctica flight~\cite{Abarr:19b}. Calculations show that the 12~m long {\it XL-Calibur} truss will satisfy the same requirement and will have no eigenfrequencies below $\sim$10~Hz. 

%
\subsection{The X-ray mirror} 
\begin{figure}[t]
\begin{center}
    \includegraphics[width=.80\linewidth]{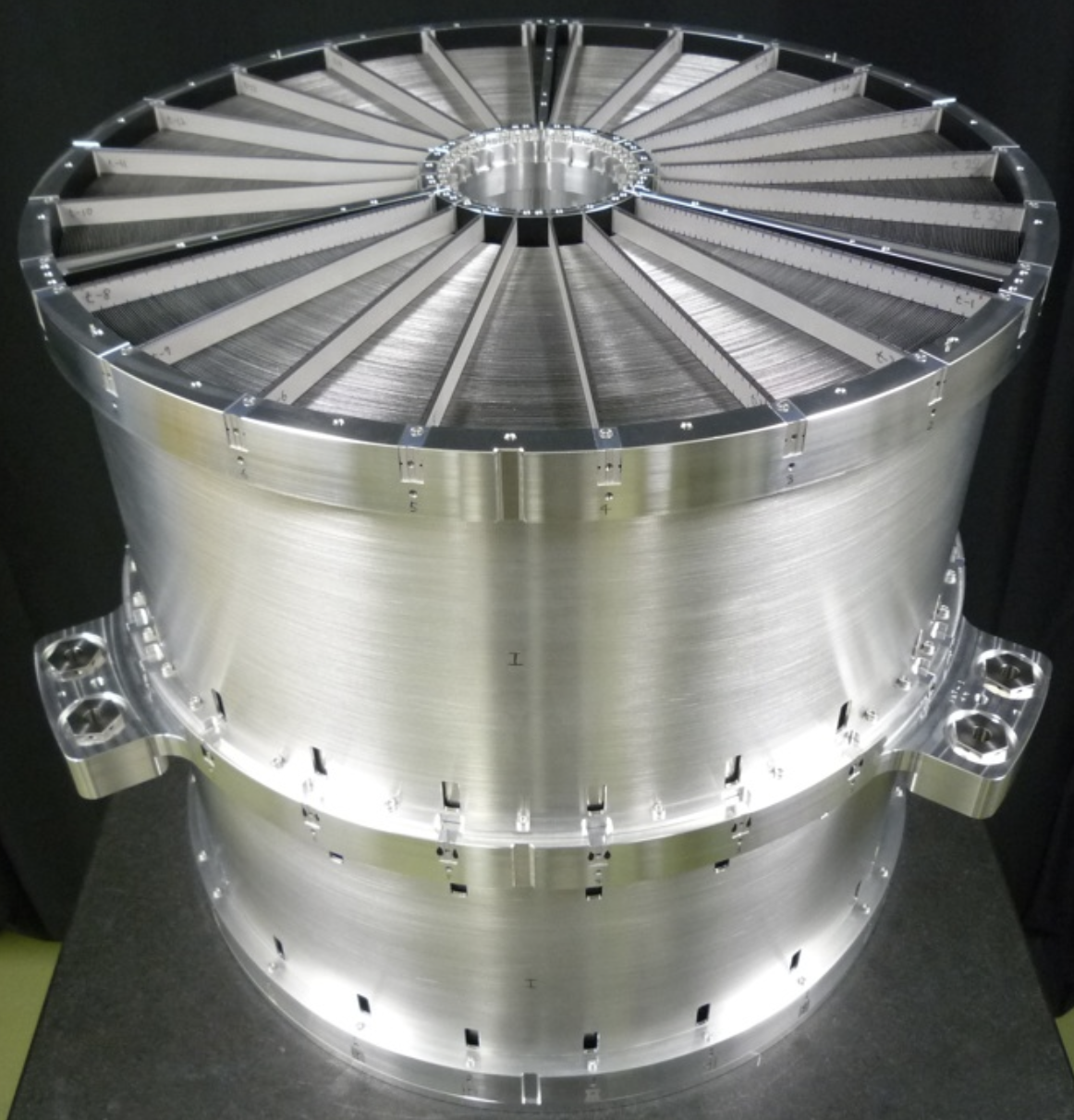}
\end{center}
\caption{\label{f:mirror} The X-ray mirror for the {\it XL-Calibur} mission (diameter 45~cm). 
The mirror was originally fabricated for the {\it FFAST} mission and is identical to that used in the {\it Hitomi} Hard X-ray Telescope mirror without the precollimator installed.}
\end{figure}

\begin{figure}[t]
\begin{center}
\includegraphics[width=.9\linewidth]{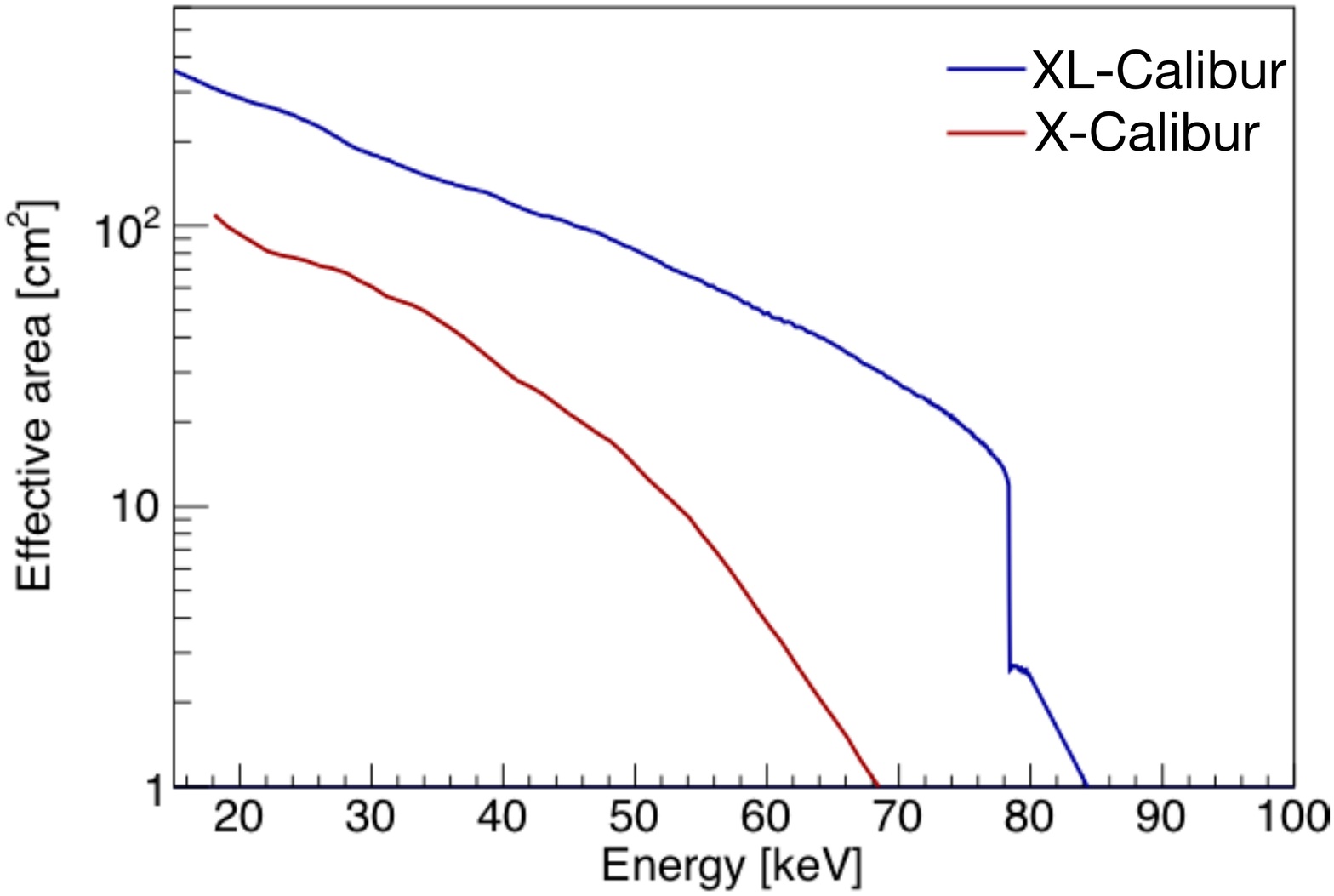}
\end{center}
\caption{\label{f:effArea} Energy-dependent effective areas of the {\it XL-Calibur} ({\it FFAST}) mirror (upper curve) and the {\it X-Calibur} ({\it InFOC$\mu$S}) mirror (lower curve)~\cite{Aeff_IF}.}
\end{figure}

{\it XL-Calibur} uses the 12~m focal length mirror originally fabricated for the {\it FFAST} mission~\cite{2014SPIE.9144E..2RT}. 
The mirror is identical to that used in the {\it Hitomi} Hard X-ray Telescope (HXT), but the precollimator is not installed. 
Since {\it XL-Calibur} will only observe bright sources along the optical axis, this does not affect the scientific performance.
A description of the mirror specifications and tests can be found in~\cite{2012SPIE.8443E..24A,2014SPIE.9144E..26A,2014ApOpt..53.7664A,2017SPIE10399E..0RA,2016SPIE.9905E..12A,2018JATIS...4a1209T,2018JATIS...4a1210M,2018JATIS...4a1212M}. 
The mirror and its energy-dependent effective area are shown in Figs.~\ref{f:mirror} and~\ref{f:effArea}, respectively.

The mirror has a diameter of 45 cm, and is made of 213 nested shells of aluminium reflectors. 
Each reflector is coated with a platinum-carbon multilayer coating with excellent grazing-incidence reflectivity from a few~keV to 80~keV. 
The mirror has an effective area of 300~cm$^2$ at 20~keV, 180~cm$^2$ at 30~keV and 130~cm$^2$ at 40~keV. 
The effective area drops at 78~keV owing to the K absorption edge of platinum.
The Half Power Diameter (HPD) of the Point Spread Function (PSF) is expected to be 2\sd after final alignment studies at the SPring-8 synchrotron facility~\cite{Aeff_IF}.

%
\subsection{Polarimeter and anticoincidence shield}
\label{ss:poldesign}
\begin{figure}[t]
\begin{center}
\includegraphics[width=.99\linewidth]{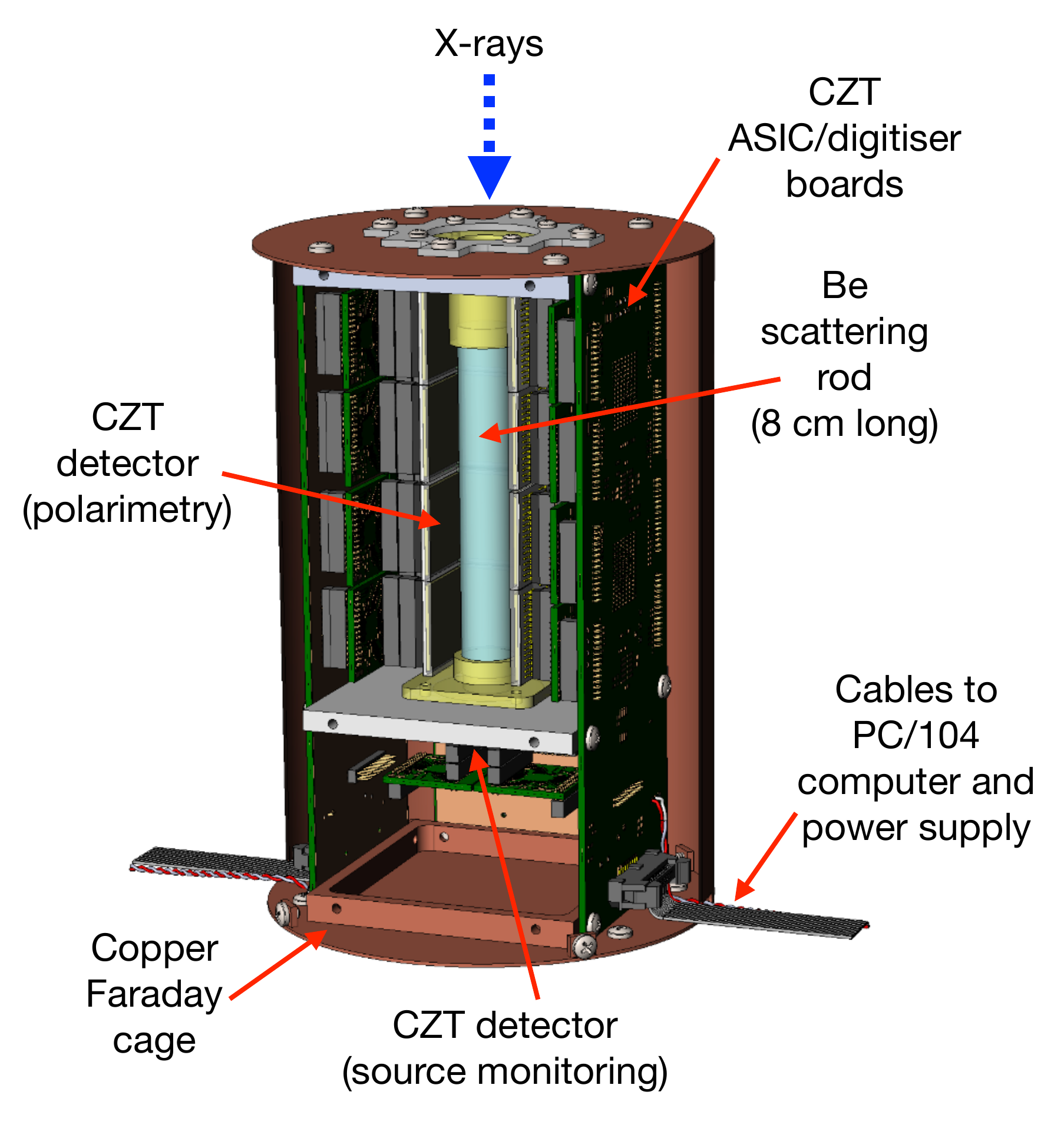}
\end{center}
\caption{\label{f:czt} A cut-away view of the {\it XL-Calibur} polarimeter.}
\end{figure}
\begin{figure}[t]
\begin{center}
\includegraphics[width=.99\linewidth]{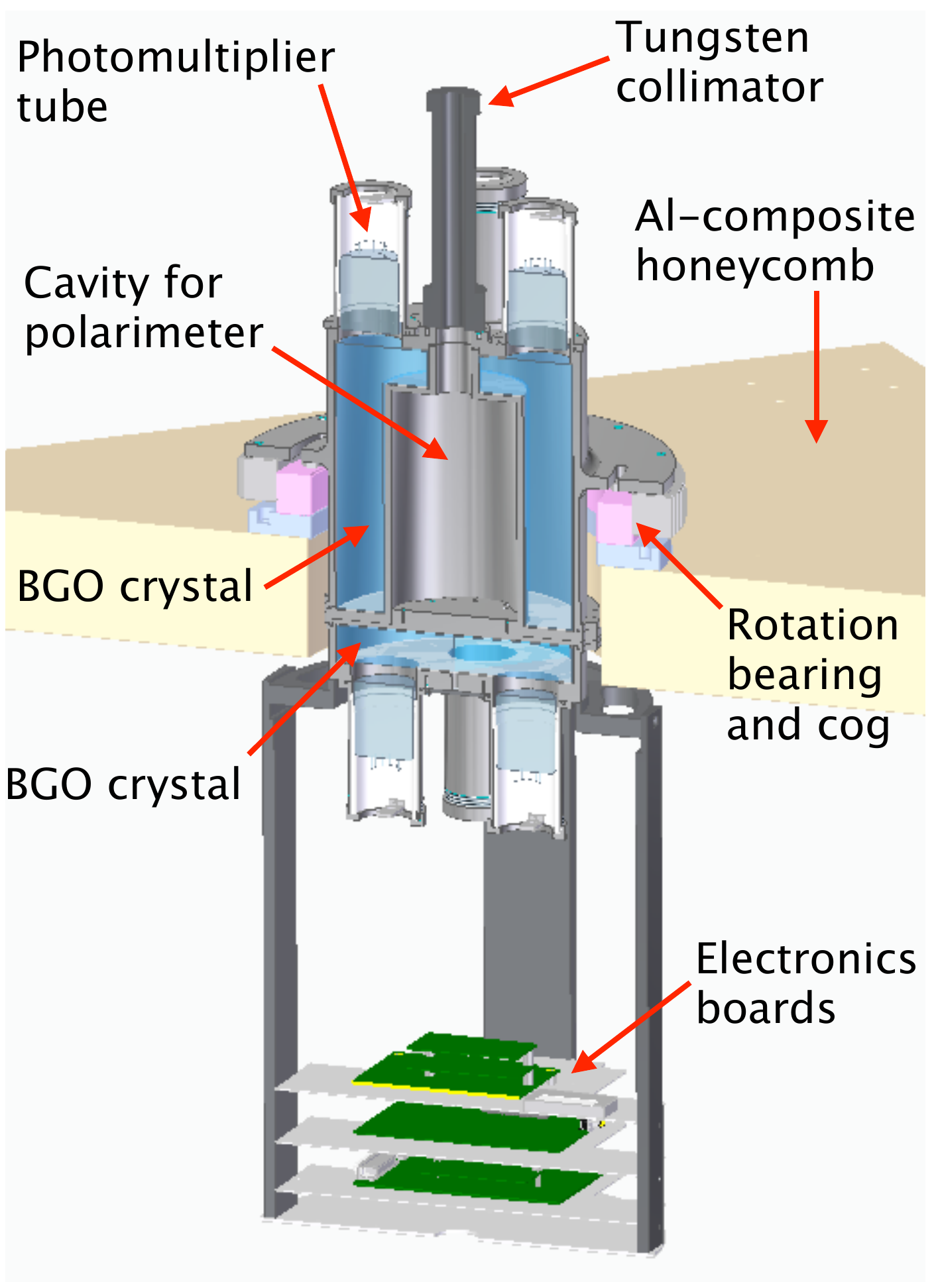}
\end{center}
\caption{\label{f:bgo} A cut-away view of the {\it XL-Calibur} anticoincidence shield mounted on the {\it X-Calibur} aluminium-composite honeycomb panel, which is mounted at the end of the truss. 
The assembly shown is $\sim$1~m long.}
\end{figure}

The design of the {\it XL-Calibur} polarimeter is shown in Fig.~\ref{f:czt}. It is closely related to the design used for {\it X-Calibur}. 
Incident photons pass through a tungsten collimator (shown in Fig.~\ref{f:bgo}) and impinge on a 1.2~cm diameter, 8~cm long, beryllium rod. The diameter is matched to the mirror PSF so that 67\% of the X-rays collected by the mirror impact the rod. 
While a larger diameter rod would intersect a higher fraction of the incoming X-rays, it would also absorb a larger fraction of the scattered X-rays. 
The beryllium rod is surrounded by 4 sets of 4 circumadjacent CZT detectors. 
The collimator prevents  direct illumination of the CZT detectors by the focussed beam.
Incident photons which do not scatter from the beryllium rod into a CZT detector may reach a 17th CZT detector, which is mounted beneath the rod. 
The signal from this detector can be used to localise the source in the field-of-view, and thus verify that the star camera/X-ray mirror/polarimeter system is correctly aligned during flight.
The polarimeter continuously rotates about the viewing axis (approximately twice per minute), which mitigates systematic effects arising from any non-uniform instrument response.

\begin{figure}[t!]
\begin{center}
    \includegraphics[width=.90\linewidth]{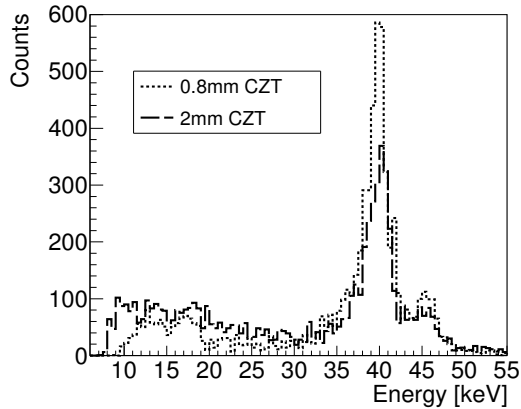}
\end{center}
\caption{\label{f:spectra} Response of 2~mm thick (dashed) and 0.8~mm thick CZT detectors (dotted) to the $\gamma$-rays of a $^{152}$Eu source. Signals from representative single pixels for each detector thickness are shown. The 2~mm and 0.8~mm thick detectors were operated at -150~V and -80~V, respectively. X-ray lines at 39.52\,keV (K$_{\alpha 2}$), 40.12\,keV (K$_{\alpha 1}$), and 45.7\,keV are evident. For the 0.8\,mm thick detector, the broader escape peaks between 10\,keV and 20\,keV are visible.}
\end{figure}

The {\it X-Calibur} 2$\times$20$\times$20 mm$^3$ CZT detectors are replaced with 0.8$\times$20$\times$20 mm$^3$ 
versions\footnote{In both cases, the detectors are provided by Kromek.} since the thinner detectors collect a factor 1.8 fewer background events.  
Both types of detector are fully efficient up to 50~keV. 
The 2~mm and 0.8~mm thick detector efficiency subsequently drops to 96\% and 74\%, respectively, at 80~keV.
The detectors are contacted with 64 anode pixels with a pitch of 2.5~mm, and a planar cathode.
Figure~\ref{f:spectra} compares $^{152}$Eu calibration results for detectors of each thickness.
The thinner detectors achieve superior energy resolution, and exhibit a smaller low-energy tail. 

The arrangement of CZT detectors and read-out electronics around the beryllium rod has been made more compact for {\it XL-Calibur}. 
This allows the inner wall of the anticoincidence shield to be located closer to the CZT detectors, thereby reducing background rates (see Section~\ref{sec:back}). 
As shown in Fig.~\ref{f:czt}, columns of four CZT detectors are arranged in a square geometry around the beryllium rod. Each CZT detector is mounted on a ceramic circuit board and interfaces with a standard circuit board containing digitising electronics based on 32-channel NRL1 Application Specific Integrated Circuits (ASIC)~\cite{Wulf} and a 12-bit analog-to-digital converter (ADC). 
The 0.8~mm thick CZT detectors read out with the NRL1 ASICs achieve a 40~keV \markp{intrinsic} energy resolution of 3.5~keV FWHM.    
Data are transmitted to a PC/104 computer via serial Low Voltage Differential Signalling (LVDS) data links.
Data are stored if one CZT pixel registers a charge deposit exceeding a configurable threshold. The recording can be inhibited if a veto signal is generated due to an energy deposit in the anticoincidence shield. 
The photon arrival time is determined with $<$5~$\mu$s accuracy using a scaler synchronised to Universal Time through the Global Positioning Satellite (GPS) system.

To mitigate particle backgrounds in the stratosphere (for further details, see Section~\ref{sec:performance}), the polarimeter is housed inside an anticoincidence shield~\cite{Iyer2020}, as shown in Fig.~\ref{f:bgo}. For {\it XL-Calibur}, BGO scintillators are used rather than the CsI(Na) used for {\it X-Calibur}.
BGO has a higher stopping power (7.1~g/cm$^2$ density, compared to 4.5~g/cm$^2$ for CsI(Na)), 
and also benefits from a faster decay time (0.3\,$\mu$s, compared to 0.46\,$\mu$s and 4.18\,$\mu$s for CsI(Na)~\cite{1969NucIM..68....9K}).
The shield comprises two parts: an inverted well, which covers the top (3~cm BGO thickness) and sides (4~cm BGO thickness) of the polarimeter, and a puck, which covers the bottom of the polarimeter (3~cm BGO thickness). The total BGO mass used in the well and puck is 35.1~kg and 6.9~kg, respectively. 

Each BGO crystal assembly is housed in a light-tight aluminium structure. The two parts of the shield are bolted together, with slots provided in the mechanical structure for routing the polarimeter cables. The distance between the BGO crystals in the two halves of the shield is 13~mm. There is no direct path to the polarimeter through this passive part of the shield. The mechanical envelope of the shield is compatible with the flight-proven {\it X-Calibur} aluminium-composite honeycomb panel, which interfaces the polarimeter-shield assembly to the truss.
Each part of the shield is read out by 4 photomultiplier tubes (Hamamatsu R6231-100) for redundancy and to ensure efficient light collection in order to achieve a 100~keV veto threshold. The lower light yield of BGO ($\sim$10~photons/keV) compared to CsI(Na) ($\sim$41~photons/keV) may be mitigated by the choice of photomultiplier gain (operating voltage).

For {\it X-Calibur}, the shield veto energy threshold was planned to be 150\,keV, but a threshold of only $\sim$1~MeV was achieved during the flight.  
This situation arose due to the passage of minimum ionising cosmic rays through the CsI(Na). The resulting large photomultiplier pulses (energy deposits of several tens of MeV) caused the shield read out electronics to saturate, producing a large dead-time ($\sim$50~$\mu$s) for each such event.
As a result, the measured background rate for {\it X-Calibur} was higher than expected.
To avoid this issue, three design changes have been implemented for {\it XL-Calibur}: 
$(i)$ the photomultiplier dynode bleeder circuit has been redesigned including clamping diodes to limit the anode signal amplitude~\cite{Tanihata_1999}; 
$(ii)$ the front-end electronics use a faster amplifier, shaper, and discriminator chain with pole-zero compensation to ensure that the system can veto a cosmic-ray rate of up to $\sim$50~kHz;
and, $(iii)$ the digital veto pulse timing and duration have been optimised, which results in a higher duty-cycle.
With these changes, laboratory tests show that a veto threshold of 100~keV is possible for a several hundred~kHz rate of large pulses, which would have saturated the {\it X-Calibur} electronics. 
The design changes have been adopted for the new BGO shield, and will also be applied to the CsI(Na) shield so that it can be used as a fall-back solution for {\it XL-Calibur} should the development of the BGO shield be delayed.
The implication for background rejection for both types of shield is discussed in Section~\ref{sec:performance}.

%
%
\subsection{Power and thermal design}
Stratospheric balloon flights from high-latitude locations are characterised by largely continuous solar illumination. 
{\it XL-Calibur} uses the same type of photovoltaic (PV) arrays as the 2018/19 {\it X-Calibur} flight. 
The PV panels come from the company SunCat Solar and use SunPower E66 solar cells laminated onto a robust honeycomb panel. 
{\it XL-Calibur} has the same power consumption as 
{\it X-Calibur} (210~W for the polarimeter and mirror heaters) and 250~W for the WASP and NASA Columbia Scientific Ballooning Facility (CSBF) components. The PV power is managed by a TriStar MPPT60 charge controller, which regulates the 24~V bus supply for the polarimeter and for the Panasonic LC-X1220AP AGM rechargeable batteries. The batteries provide 4--6 hours of back-up power. 

During the planned balloon flights, the gondola will be illuminated by continuous but variable sunlight.
Maintaining thermal control of the payload is an important aspect of minimising systematic effects during measurements, e.g.\,to ensure that the polarimeter has a  uniform response, and that the truss does not deform due to differential heating effects.
In order to ensure predictable thermal behaviour, it is common practice to cover surfaces exposed to solar radiation in reflective  aluminised mylar sheets, and white teflon tape. 
Data from the previous {\it X-Calibur} flight~\cite{Abarr:19b} have been used to assess the thermal modelling approach and inform the {\it XL-Calibur} thermal design.
For electronic components, the {\it XL-Calibur} thermal design approach is driven by the low ambient pressure at float altitude, which means that the primary heat transfer mechanisms are radiation and conduction. 
Dedicated heat-conduction paths are established between high power-dissipation components and radiating mechanical enclosures. 
All active components will be tested from $-$30\,$^{\circ}$C to +50\,$^{\circ}$C, as well as being tested at the low pressure present at flight altitude. 
           
%
\subsection{Preflight calibration and alignment procedures}
The positive-definite nature of polarimetric measurements requires that both unpolarised and polarised X-ray beams are used when characterising the polarimeter response.
Highly-collimated high-rate X-ray beams can be produced at synchrotron facilities across a range of energies. 
Beams with $\sim$100\% linear polarisation can be delivered by scattering a primary beam off a crystalline material.
The {\it X-Calibur} polarimeter was characterised at the Cornell High Energy Synchrotron Source~\cite{2014JAI.....340008B}. 
In the laboratory, or at the launch site, a beam with $\sim$100\% linear polarisation may be formed using a radioactive source, e.g. $^{241}$Am, by scattering its X-ray beam (59.5~keV) through 90$^\circ$~\cite{Chauvin:2015qda}. Unscattered beams from radioactive sources can also be used to confirm the energy response of the polarimeter and the shield in the field. 

A number of alignment studies are required to control systematic effects, which arise when measuring polarisation and to ensure that the photon detection efficiency is maximised during flight. 
The optical axes of the X-ray mirror and polarimeter must be aligned to ensure that focussed X-rays impinge on the centre of the beryllium scattering rod. An alignment procedure has been developed, which can be used at the launch site, based on a collimated beam of visible light, which reflects in the mirror identically to X-rays. 
The set-up comprises a laser diode placed at the eyepiece of a 356~mm (14~inch) diameter Celestron Schmidt-Cassesgrain telescope to produce a virtual light source at infinity, thereby allowing a parallel beam of visible light to enter the X-ray mirror. During the alignment process, the incident direction of the light beam is recorded, and the location of the focal spot in the detector plane is monitored using two cameras (``forward-looking'' and ``backward-looking''), which are permanently mounted inside the X-ray mirror on the optical axis. 
During the {\it X-Calibur} flights the orientation of the forward-looking camera and the WASP star camera were cross-calibrated using star images.
The truss bending was determined with a precision of 0.1~mm using the backward-looking camera to survey an LED target mounted on the polarimeter collimator~\cite{Abarr:19b}. 
The procedure allows the focal plane position of any target in the star camera field-of-view to be determined accurately. 
Additionally, standard metrology techniques (a theodolite-mounted laser and telescope system surveying alignment cubes mounted on payload components) are used to co-align the bore-sights of the two star cameras, to co-align the star cameras to the sun sensor, and to co-align the star camera to the X-ray axis of the polarimeter.
The polarimeter rotation angle is also aligned relative to the star tracker bank angle.

%
%
\section{Design Optimisation and Estimated Performance}
\label{sec:performance}
The {\it XL-Calibur} design has been studied and optimised using a Monte Carlo approach implemented with the Geant4~\cite{2003NIMPA.506..250A,2016NIMPA.835..186A} simulation package (version 10.04p03). 
The simulation geometry includes the beryllium scattering rod, CZT detectors, copper Faraday cage surrounding the polarimeter, aluminium-encased BGO anticoincidence shield, tungsten collimator, and rotation bearing assembly. This provides a realistic representation of the material distribution in the vicinity of the CZT detectors. 
The interactions of both source photons and background particles with the simulation model volumes have been considered\footnote{The Geant4 'shielding' physics list~\cite{Koi:10} modified with the 'Livermore polarisation' physics list is used.}.
For incident particles interacting with the CZT or the anticoincidence shield, the interaction location and deposited energy are stored.
The CZT energy deposits are converted to measured ADC channels according to an experimentally determined response function. 
The conversion has been tested using data from the {\it X-Calibur} 2018/2019 flight and good agreement is found between the simulated and observed energy spectra~\cite{Abarr:19b}.

%
\subsection{Signal}
\begin{figure*}[t!]
    \centering
    \includegraphics[width=0.7\linewidth]{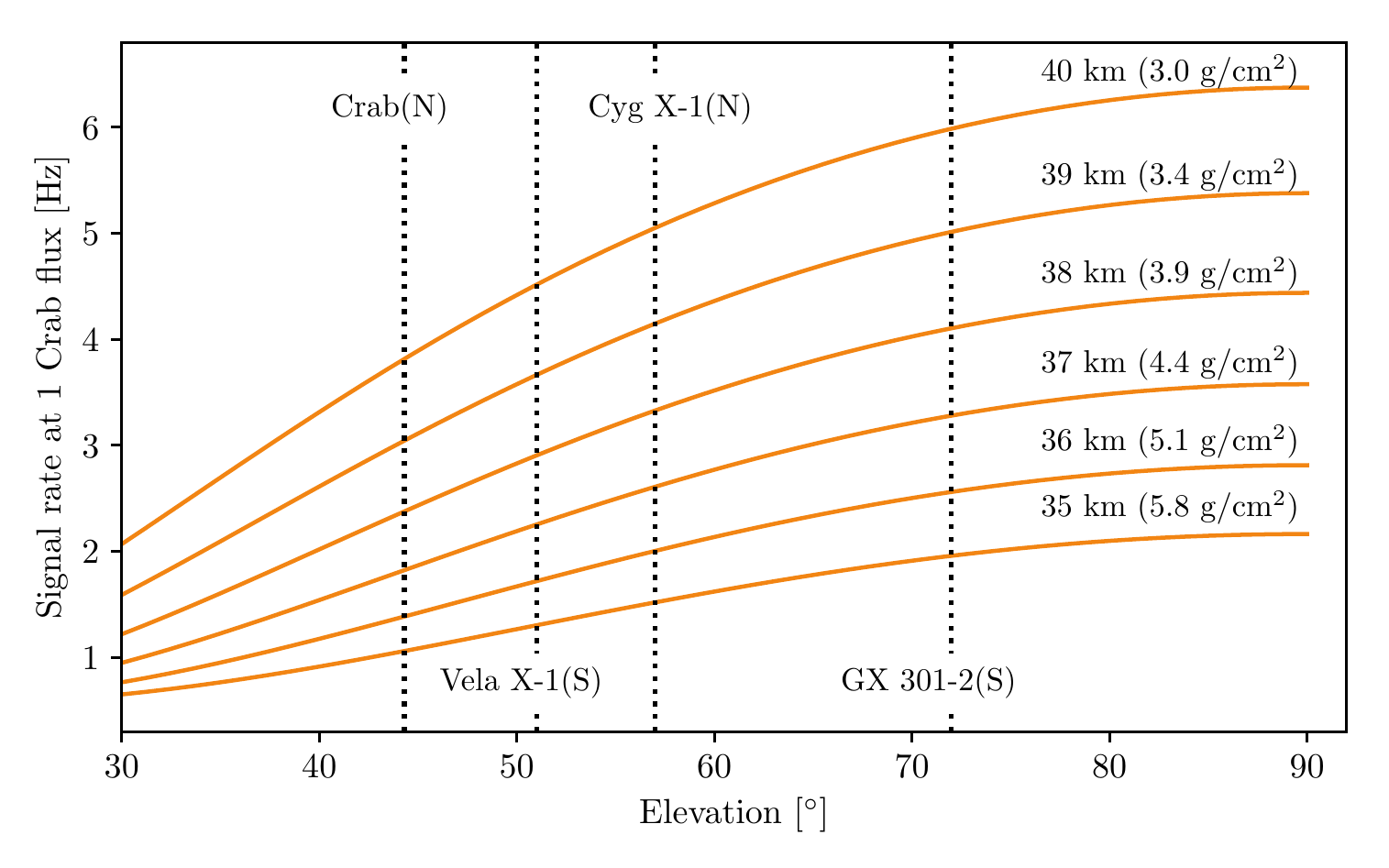}
    \caption{The variation in signal rate for a 1~Crab source as a function of source elevation. The effect of observing altitude is shown. The vertical lines denote the maximum elevation of potential sources during balloon flights from either Esrange (N) or Antarctica (S).}
    \label{fig:sig}
\end{figure*}

The signal response was determined using a beam of photons directed through the collimator. 
Photon energies ($E$) were distributed according to the Crab energy spectrum,   
\mbox{$F(E)=9.42E^{-2.12}$~photons keV$^{-1}$ cm$^{-2}$ s$^{-1}$}~\cite{Kirsch_2005}.
In Fig.~\ref{fig:sig}, the signal rate is shown as a function of source elevation, where the effect of observing altitude is included. 
For a representative observing altitude of 38~km, the average signal rate for a 1~Crab source varies between $\sim$2--4.4~Hz in the elevation range 40--80$^\circ$. 

\begin{figure*}[t!]
    \centering
    \includegraphics[width=0.7\linewidth]{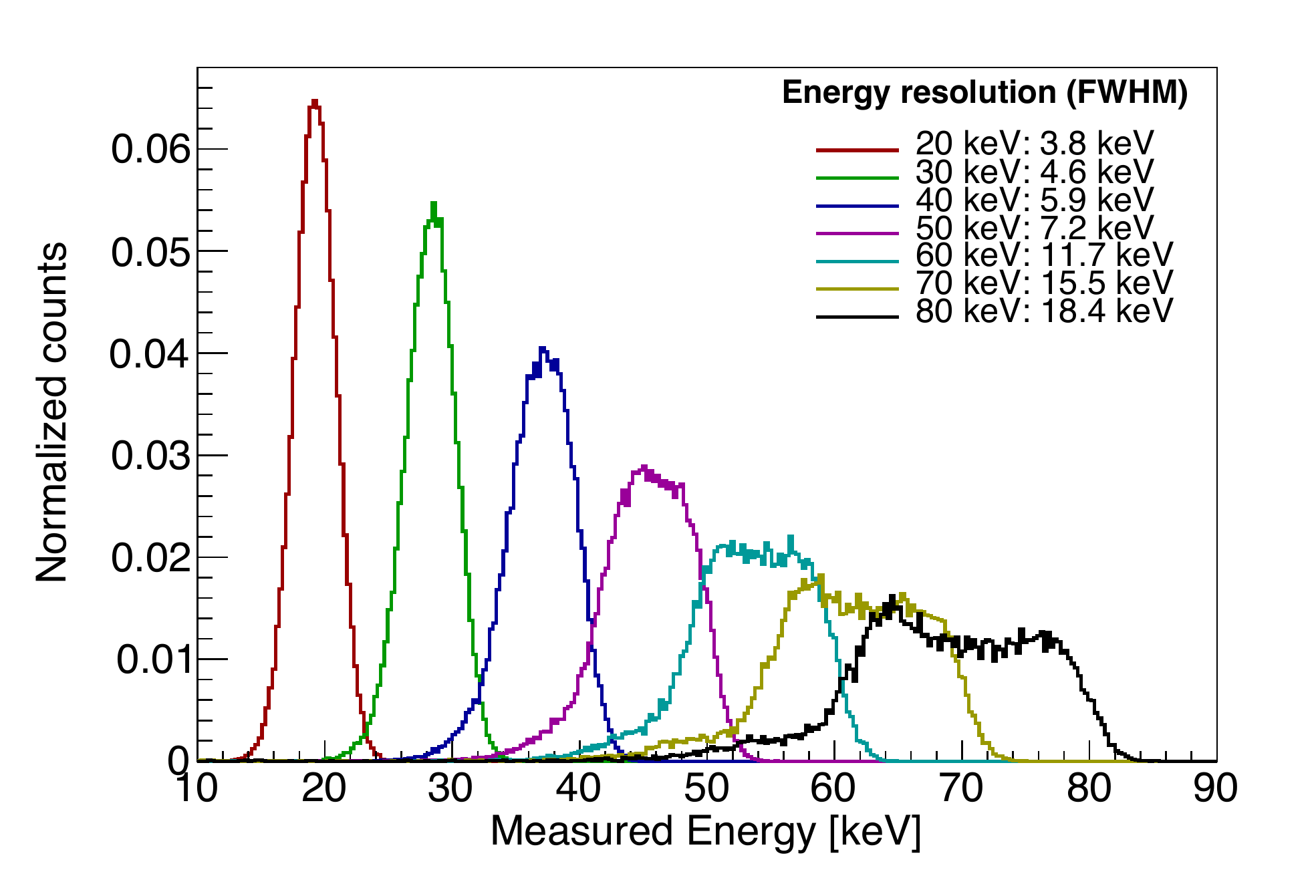}
    \caption{\markp{The simulated energy resolution for a range of mono-energetic photons which scatter off the beryllium rod before impinging on one of the 16 circumadjacent CZT detectors. An energy-independent intrinsic CZT detector energy resolution of 3.5~keV is assumed.}}
    \label{fig:czt_res}
\end{figure*}

\markp{The energy resolution when combining signals from all 16 CZT detectors has been simulated in the {\it XL-Calibur} energy range. In contrast to the result shown in Fig~\ref{f:spectra}, the simulation accounts for the energy loss of photons scattering from the beryllium rod and demonstrates the effect of Compton scattering energy losses within the CZT detectors. 
As shown in Fig.~\ref{fig:czt_res}, the energy resolution worsens with energy. At 40~keV the measured energy resolution is 5.9~keV, compared to the intrinsic CZT detector energy resolution of 3.5~keV.
Studies are in progress to determine if it is possible to improve the energy resolution at high energies through selections on the polar scattering angle.}

%
\subsection{Background}
\label{sec:back}
The energy spectra of background particles at a specified atmospheric depth are produced for a given primary flux of cosmic-ray protons and helium nuclei incident on the top of the atmosphere using the MAIRE code\footnote{Models for Atmospheric Ionising Radiation Effects, {\tt http://www.radmod.co.uk/maire}. This code was previously known as QARM~\cite{2006ITNS...53.1851L}.}. 
The primary flux depends on the observing position (latitude, longitude), geomagnetic activity (Kp-index) and the observation date (amount of solar modulation). 
A balloon altitude of 38~km is considered for solar minimum and solar maximum conditions with a low geomagnetic activity index (Kp=2). 
MAIRE generates identical energy spectra for background particles at Antarctica (latitude: $-77.84^\circ$, longitude: $166.68^\circ$) and Esrange (latitude: $67.86^\circ$, longitude: $20.23^\circ$). This is because MAIRE only considers a hadronic primary cosmic-ray flux, which is relatively insensitive to the difference in rigidity cut-off at these locations.
The primary flux of cosmic-ray electrons is more affected by geomagnetic location, but the flux is two orders of magnitude lower than the hadronic flux~\cite{pamela} and the resulting background is not important for {\it XL-Calibur}. 

\begin{figure*}[t!]
    \centering
    \includegraphics[width=0.7\linewidth]{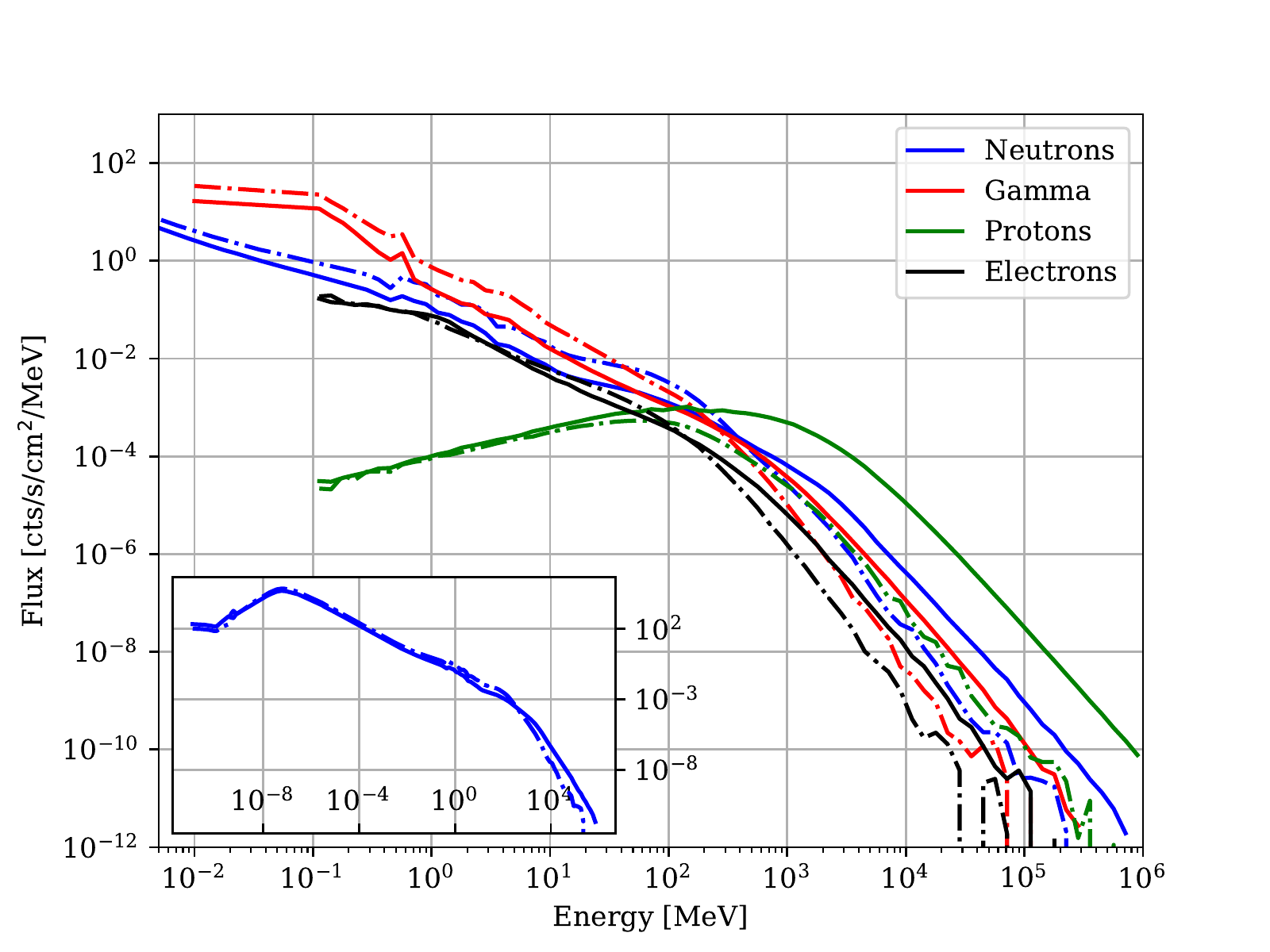}
    \caption{The energy spectra of background components generated by the MAIRE simulation. Up-going (down-going) fluxes are shown as solid (dashed) lines. The inset figure shows the neutron spectrum, which covers a wider energy range than the other components.}
    \label{fig:background}
\end{figure*}

The resulting background-spectra inputs to Geant4 comprise up- and down-going atmospheric electrons, neutrons and photons, up-going atmospheric protons, and down-going atmospheric and primary protons. 
After attenuation by the atmosphere, the contribution from cosmic X-ray background photons~\cite{CXB} is approximately two orders of magnitude lower than secondary X-/gamma-rays produced in the atmosphere.
The energy spectra for all the simulated background components are shown in Fig.~\ref{fig:background}.

A source photon, which enters the beryllium scattering rod and subsequently scatters into one of the surrounding CZT detectors, constitutes a signal event in {\it XL-Calibur} if there is no coincident anticoincidence veto signal.  
Despite the thick anticoincidence shield and passive materials surrounding the polarimeter assembly, background particles can produce an identical signature.
This background could be largely eliminated by replacing the beryllium rod with an active plastic scintillator and requiring temporal coincidence between the scintillator and CZT signals~\cite{2014JAI.....340008B}. 
A significant drawback with this approach is that a large fraction of the Compton scattered events in the {\it XL-Calibur} energy range deposit only a few keV in a plastic scattering element, resulting in a $\sim$50\% detection efficiency. 
The beryllium scattering rod has a lower atomic number than plastic scintillator, yielding a higher scattering-to-photoelectric cross-section ratio~\cite{2018JATIS...4a1004K}. Moreover, the higher density of beryllium means that photons are more likely to scatter near the top of the beryllium rod, which enhances the signal-to-background ratio in the upper rings of CZT detectors.

\begin{figure}[t!]
    \begin{center}
    \includegraphics[width=\linewidth]{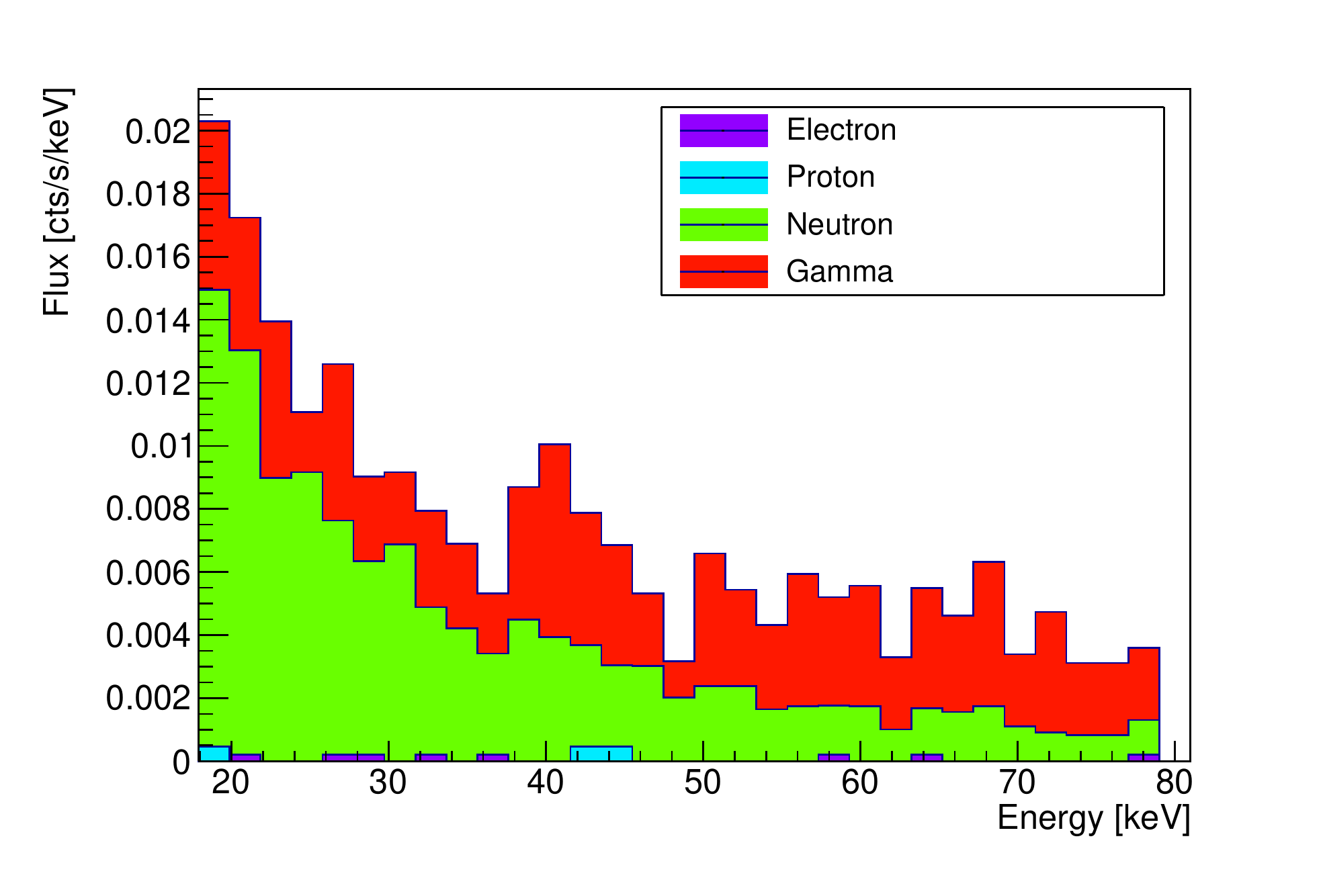}
    \end{center}
    \caption{Stacked histograms showing the components of the CZT background spectrum for {\it XL-Calibur} equipped with a BGO anticoincidence shield. For comparison, the flux presented by a 200~mCrab source is $\sim$0.025~counts s$^{-1}$ keV$^{-1}$ at 30~keV.}
    \label{fig:relcont}
\end{figure}

As introduced in Section~\ref{ss:poldesign}, several approaches are being adopted to reduce the occurrence of background events compared to {\it X-Calibur}: 
{\it (i)} a high stopping power BGO anticoincidence shield;
{\it (ii)} a more compact polarimeter assembly, allowing the shield inner wall to lie closer to the CZT detectors;
{\it (iii)} a significant reduction of the shield veto energy threshold;
{\it (iv)} reducing the thickness of the CZT detectors from 2\,mm to 0.8\,mm.
The effect of these measures is shown in Table~\ref{tab:bgrates} and can be summarised as follows:
\begin{enumerate}
\item Decreasing the anticoincidence veto threshold to $\sim$100~keV reduces the background rate by a factor of 2.6 compared to {\it X-Calibur}. 
\item Reducing the CZT thickness from 2\,mm to 0.8\,mm reduces the background rate by an additional factor of 1.8.
\item Implementing a more compact polarimeter assembly and BGO anticoincidence shield further reduces the background rate by a factor of 5.4.  
\end{enumerate}
Overall, the background rate is reduced by a factor $\sim$25. 
Figure~\ref{fig:relcont} shows the composition of the background as a function of the energy deposited in the CZT detectors. 
The background is dominated primarily by (mainly albedo) atmospheric $>$100~MeV neutrons, which penetrate the anticoincidence shield, as well as 
$\sim$MeV atmospheric gamma-rays. The effect of surrounding the anticoincidence shield with a polyethylene neutron moderator~\cite{2016APh....82...99C} of thickness 5~cm (8~cm in the vicinity of the CZT detectors) has been studied.
The background is reduced by a factor of 1.2, but the moderator increases the polarimeter mass by at least 30~kg, which potentially decreases the observation altitude and places more complex requirements on the mechanical design. The moderator is therefore not implemented.  

\begin{table*}[t!]
    \centering
   \begin{tabular}{|l|c|c|}
        \hline
         {\bf Configuration} & \multicolumn{2}{|c|}{\bf Background rate (Hz)}\\      
          & E$_{\mathrm{th}}$=100~keV & E$_{\mathrm{th}}$=1~MeV \\
         \hline
         {2~mm CZT + CsI ({\it X-Calibur})} & 4.93$\pm$0.10 & 12.64$\pm$0.16$^\dagger$ \\
         0.8~mm CZT + CsI & 2.74$\pm$0.08 & 7.36$\pm$0.13 \\
         0.8~mm CZT$^*$ + BGO & 0.51$\pm$0.02$^\mathsection$ & 2.24$\pm$0.04 \\
         0.8~mm CZT$^*$ + BGO + polyethylene & 0.44$\pm$0.01 & 2.01$\pm$0.04 \\         
         \hline
    \end{tabular}
    \caption{Simulated background rates (15-80~keV, Hz) for different polarimeter design configurations assuming 2018/2019 {\it X-Calibur} flight conditions. Two anticoincidence energy thresholds, E$_{\mathrm{th}}$, are considered. Statistical errors (simulation statistics) are quoted.  
    \footnotesize{$^\dagger$A lower background (2.3~Hz) was reported in~\cite{Abarr_2020} because a narrower energy range was considered (15-35~keV), and not all CZT rows were used. $^*$With compact CZT configuration. $^\mathsection$Predicted for {\it XL-Calibur.}}}
    \label{tab:bgrates}
\end{table*}

%
\subsection{Systematic errors} 
During detailed studies of the systematic errors for {\it X-Calibur}~\cite{2014JAI.....340008B}, two factors were found to dominate:
\begin{itemize}
\item Knowledge of the modulation factor. For {\it X-Calibur}, the modulation factor was measured with a relative accuracy of $<2\%$. 
This resulted in a relative error on the measured polarisation fraction $p_{\rm r}$ of $0.02 p_{\rm r}$. 
\item During observations, the {\it XL-Calibur} pointing direction will be varied between the source location (on-source observations) and background fields with a $\sim$1$^{\circ}$ degree offset (off-source observations). 
{\it X-Calibur} adopted this approach and could demonstrate null polarisation for background observations~\cite{Abarr:19b}. 
Incorrectly estimating a time-variable background would lead to an under- or over-subtraction of the background. 
This may increase or decrease the measured polarisation fraction, respectively. The importance of this effect will depend on the signal-to-background ratio and will be more important for dim sources than for bright sources. 
\end{itemize}
Alignment systematics (e.g. a mirror-detector misalignment) produce a dominant dipole contribution in angular space as the polarimeter rotates about the viewing axis. The 360$^{\circ}$ periodicity in the scattering angle compared to a period of 180$^{\circ}$ for the polarisation signal allows the two effects to be disentangled. 

%
%
\section{Science Programme}
\label{s:science}
The science return from the foreseen observing programme is presented in this section.
Since the X-ray sky is variable, the observation programme
will be optimised prior to and during the flights based on visibility
constraints, and the fluxes measured by the X-ray and $\gamma$-ray
missions available at the time. For the {\it X-Calibur} flight,
monitoring data from {\it Swift} BAT\footnote{\tt
https://swift.gsfc.nasa.gov/results/transients/ BAT\_current.html} was
used.

A 345~ksec integration time is anticipated for flights from Esrange to
Canada (5 days, 80\% efficiency), and between 552~ksec and 3.8~Msec for
a circumpolar McMurdo flight (8--55 days, 80\% efficiency). For an
Esrange flight, at least two targets will be observed extensively (e.g.
Cyg\,X$-$1 and Her\,X$-$1) and two other targets with shorter exposures
(including the bright Crab pulsar). For a longer McMurdo flight,
approximately 4--10 targets can be observed at flux levels exceeding 150~mCrab. Background levels are conservatively taken to be a factor of 10
lower than for the 2018/2019 {\it X-Calibur} flight. Launch date
constraints may preclude the observation of some targets (e.g. for an
Esrange flight, the Crab can be observed during the long-duration flight
window in May, July and August, but is too close to the Sun in June).
For {\it X-Calibur}, 50\% of the observing time was spent observing
off-source. {\it XL-Calibur}'s improved signal-to-background ratio
allows the fraction of off-source pointings to be lowered to $\sim$35\% of
the total integration time~\cite{2015APh....68...45K}.
%
\subsection{Revealing the geometry and location of the X-ray bright corona of accreting stellar-mass black holes}
\begin{figure}[t!]
\begin{center}
\includegraphics[width=\linewidth]{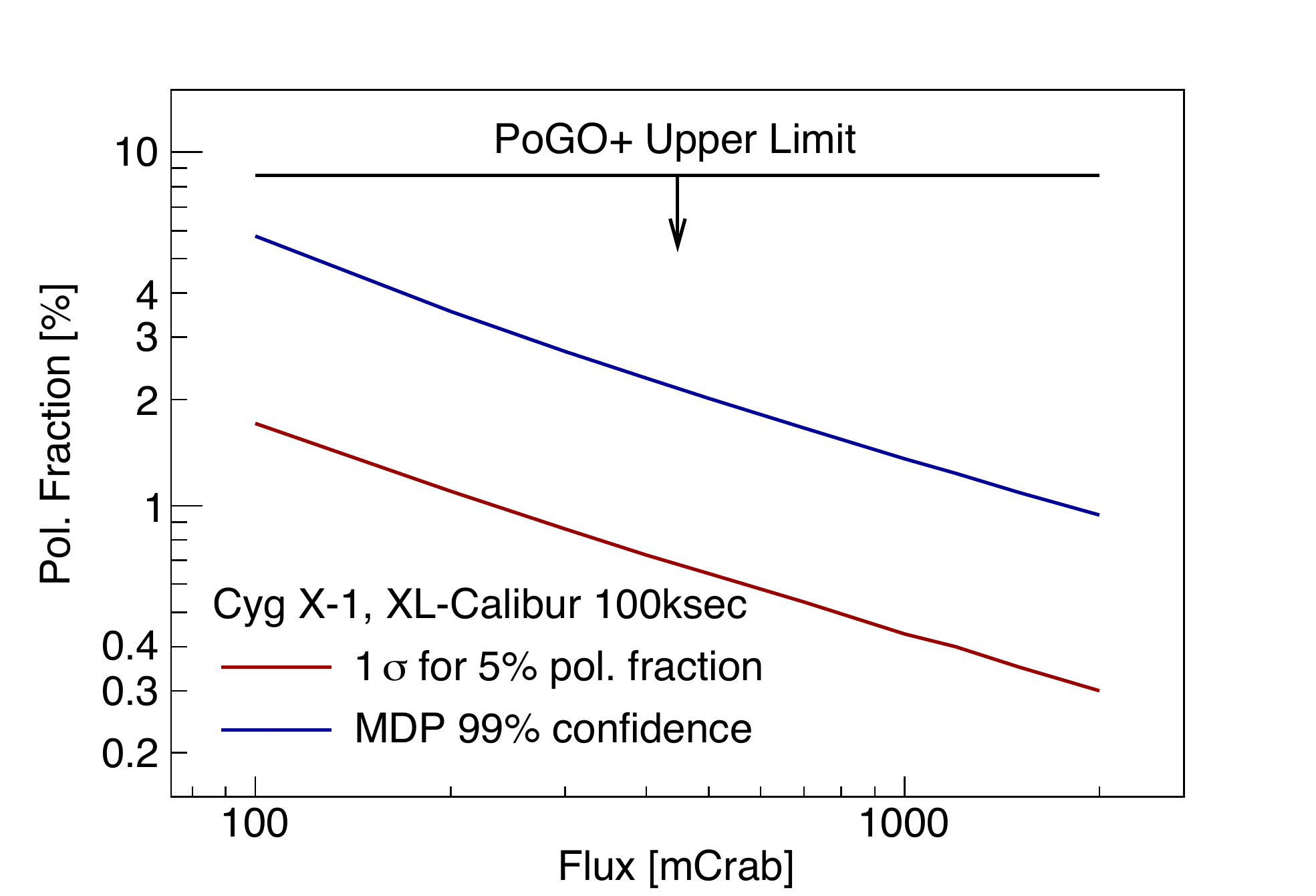}
\end{center}
\caption{\label{f:cyg} The {\it XL-Calibur}
observations of Cyg\,X$-$1 will improve substantially over the {\it PoGO+} observations. 
For 15-45 keV flux levels between 200 mCrab and 2 Crab,
a 100~ksec on-source observation by {\it XL-Calibur}
achieves 99\% confidence level MDPs between 4\% and 1\% (blue line)
and statistical 1$\sigma$ polarisation fraction errors between
1\% and 0.3\% (red line). An atmospheric column density of 5~g/cm$^2$ is assumed.}
\end{figure}

During a week-long balloon flight from Esrange to Canada in 2016, 
{\it PoGO+} observations constrained the polarisation of Cyg\,X$-$1 emission
(19--181~keV) in the low-hard state to be less than 8.6\% (90\%
confidence level)~\cite{2018NatAs...2..652C}. {This value is
commensurate with high scattering opacity accretion disks when viewed at
significant angles to their surface normal directions~\cite{ST85}.
Similar polarisation fractions of a few percent are produced by fully
general relativistic modeling~\cite{Dovciak-2008-MNRAS}, and also by
extended corona~\cite{2010ApJ...712..908S}.  Interestingly, the {\it PoGO+} value} is significantly lower than the 15\% model prediction of
the lamp post
model~\cite{2012JPhCS.372a2056D,2014xru..confE.243D,2017rdeh.confE..29D}.
{The lamp post geometry~\cite{Matt-1991-AandA} places a small
X-ray emitting corona at some height above the  
accretion disk, which is considered to be axisymmetric.
This defines a preferred direction, which naturally produces a
higher polarisation fraction. Usually the corona is presumed to lie
along the jet axis.} During a day-long on-source observation of
Cyg\,X$-$1 in the low-hard state, {\it XL-Calibur} will achieve a MDP of
2\% for a typical low-hard flux of 700 mCrab (Fig.\ \ref{f:cyg}).

{For a uniform, planar disk, the polarisation vector direction
(defining the polarisation angle) is perpendicular to the line that
forms the intersection of the disk surface and the plane of the sky.
Thus, it nominally lies parallel to the sky-projected direction of a jet.} The
geometrical {\it XL-Calibur} constraints can be compared with those from
VLBI observations of the jet~\cite{2001MNRAS.327.1273S} and from
observational constraints on the orbital plane~\cite{2011ApJ...742...83R}.
Cyg\,X$-$1 is occasionally observed in the high-soft state with
100--150~mCrab fluxes in hard X-rays. In this case, the corona shape can
be constrained in the high-soft state for the first time. A compact
corona within a few gravitational radii from the black hole would create
a polarisation fraction of $\sim$15\%, which {\it XL-Calibur} can detect
with high significance even at low flux levels. {Even lower,
disk-like values of $\sim$5\% will be accessible to {\it XL-Calibur}.}
The polarisation fraction and angle can be used to constrain the black hole
spin~\cite{2012JPhCS.372a2056D,2014xru..confE.243D,2017rdeh.confE..29D}.
It will be important to complement the Cyg\,X$-$1 results with
observations of other systems, e.g. GX\,339$-$4, 4U\,1630$-$40 and other
transient accreting black hole systems. If all these systems show low
polarisation fractions, generalised conclusions about the corona shape
can be made.
%

\subsection{Pin-pointing the origin of X-rays from accreting pulsars, and exploring the fundamental physics of the QED and plasma birefringence}
The neutron stars in Her\,X$-$1, \gx and Vela\,X$-$1 are prototypical mass-accreting and strongly magnetised pulsars in high-mass X-ray binaries
(HMXBs). Despite years of multi-wavelength observations, it is still not
known where and how the X-rays originate. {\it XL-Calibur} observations
will provide qualitatively new geometrical information and stand to
provide a real breakthrough in this line of research.  All three sources
are bright with average fluxes around 250 mCrab and brighten regularly
with orbital periods of 41.5 days (\gx) and 8.9 days (Vela\,X$-$1), and
a super-orbital period of 35 days (Her\,X$-$1). {The polarisation
detection by {\it X-Calibur} of \gx demonstrates the polarimetry
prospects for this source class~\cite{Abarr_2020}.   Detailed radiation
transport calculations have been performed for strongly magnetised X-ray
pulsars~\cite{Meszaros-1980-ApJ,Kaminker-1982-APSS,1988ApJ...324.1056M},
indicating 
polarisation signatures that are strongly dependent on the photon energy and its propagation angle relative 
to the field direction~\cite{Kii-1987-PASJ}. The emergent linear
polarisation therefore depends strongly on the pulsar phase and on
photon energy, especially so in the environs of cyclotron absorption
features.  It carries a clear imprint of the X-ray emission region
geometry.  The two main competing models for accreting X-ray pulsars
focus on the radiation beam shape (e.g.~\cite{Schoenherr-2007-AandA}).
For low accretion rates, emission from plasma columns, which are shocked
above the neutron star surface radiate a {pencil beam} along the
accretion flow. This contrasts with high accretion luminosity systems, which
dissipate their energy at the stellar surface and radiate a broad {fan beam} oriented with the surface. Through phase-resolved polarimetry,
since these geometries present opposite correlations between intensity
and polarisation fraction in a pulse profile~\cite{1988ApJ...324.1056M},
{\it XL-Calibur} will help discern between these two leading geometrical
pictures for these bright HMXBs.}

{{\it XL-Calibur} also affords the opportunity to study the
birefringent properties of the magnetized QED vacuum and accretion
column plasma. Polarisation of the QED vacuum by the magnetic field
leads to birefringent propagation of light~\cite{Heisenberg-1936-ZPhys,Adler-1971-AnPhy}, an effect which defines
elliptical polarisation eigenstates stemming from a refractive index, which scales as the square of the magnetic field strength.  These differ from the eigenmodes of plasma
polarisation, with plasma dispersion scaling with the square
of the plasma frequency, and therefore linearly with the
plasma density.  The competition between the two dispersive
influences generates a so-called {vacuum resonance} frequency, about
which the polarisation properties change dramatically (see~\cite{Harding-Lai-2006-RPPh} and references therein). This frequency is
thus dependent on the density and the field strength, and can naturally
fall in the hard X-ray window, enabling the prospect for {\it
XL-Calibur} to provide the first evidence for this signature
prediction of QED. Phase and energy-resolved {\it XL-Calibur}
observations in the 15--80 keV band will cover the cyclotron absorption
features of Her\,X$-$1, \gx and Vela\,X$-$1~\cite{2019A&A...622A..61S}.
The cyclotron band is rich with diagnostic potential given that the
polarisation signatures rapidly change with energy~\cite{Meszaros-1992-book} due to the physics of normal mode propagation
and the resonant interactions of light with the magnetized electrons.}

\begin{figure}[t!]
\begin{center}
\includegraphics[width=\linewidth]{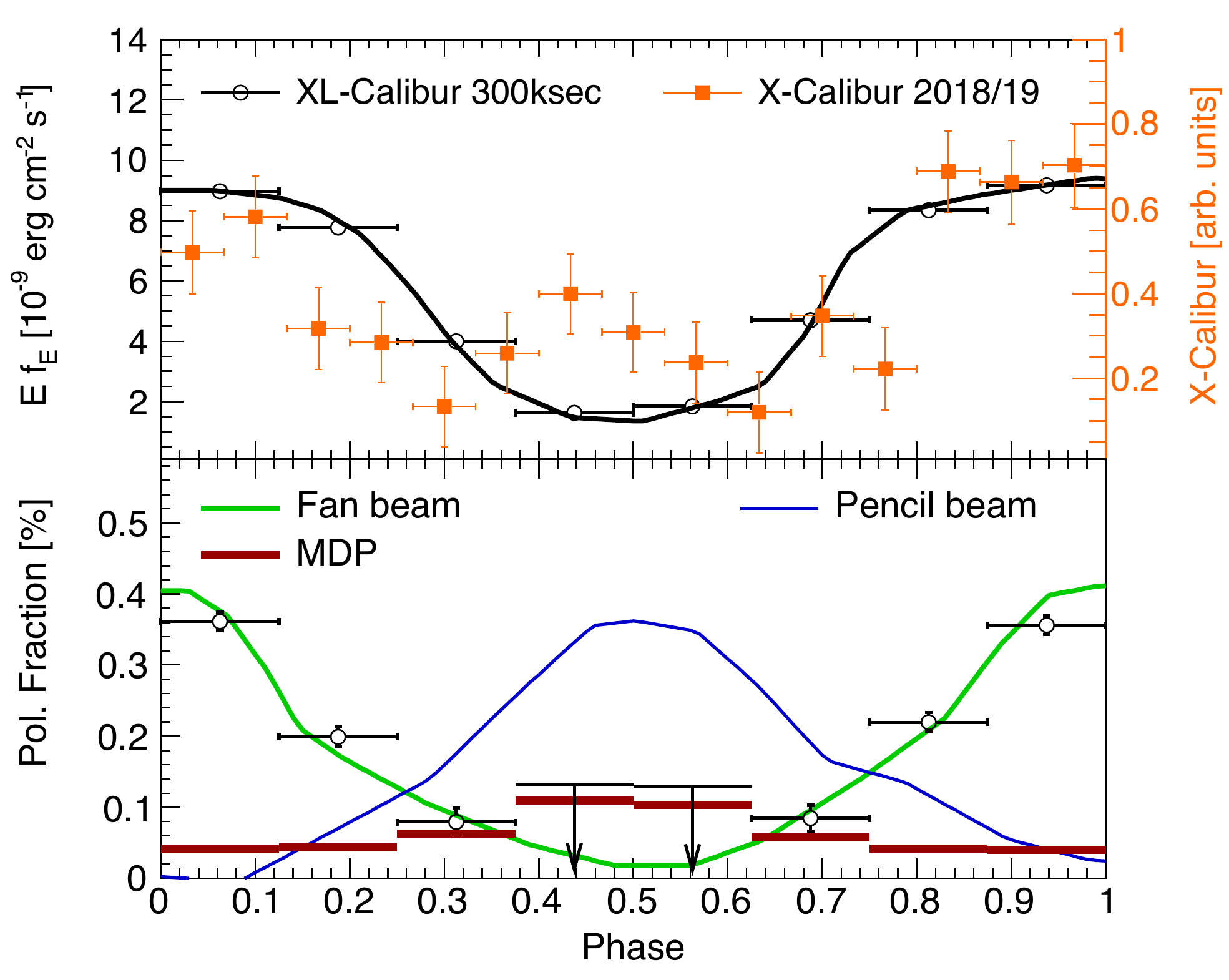}
\end{center}
\caption{\label{fig:gx301}Simulated 300~ksec \gx observation with {\it XL-Calibur} at an atmospheric column density of 7~g/cm$^2$ (the mean depth of the 2018/2019 {\it X-Calibur} flight). An energy spectrum similar to~\cite{F_rst_2018} is used with a 25-70~keV flux of 700~mCrab.
The upper panel shows the assumed pulse profile (black line), measured {\it X-Calibur} 2018/19 pulse profile (orange data points), and simulated {\it XL-Calibur} results (black data points). The lower panel shows the expected polarisation fractions for the fan beam (green line) and pencil beam (blue line) models of~\cite{1988ApJ...324.1056M} (model 45/45). The black data points show the simulated {\it XL-Calibur} polarisation fraction results for the fan beam model, and the dark red lines show the MDP values.}
\end{figure}

The expected performance for \gx is shown in Fig.~\ref{fig:gx301}. This
is an updated version of the result presented in~\cite{Abarr_2020}, with
a more realistic background level used. The simulations show that {\it
XL-Calibur} can clearly distinguish between the predictions of the fan
beam and pencil beam models of~\cite{1988ApJ...324.1056M}. The
conclusion will be largely independent of the viewing angle and the
angle between the magnetic field axis and the rotation axis of the
pulsar. The intrinsic radiation pattern is directly related to the
extent of the emitting region and the optical depth of the accretion
column. The information will also be important to explain variations in
the cyclotron absorption energies~\cite[e.g.][]{2019A&A...622A..61S}.
%

\subsection{Observations of the Crab nebula and pulsar}
\begin{figure}[t!]
\begin{center}
   \includegraphics[width=\linewidth]{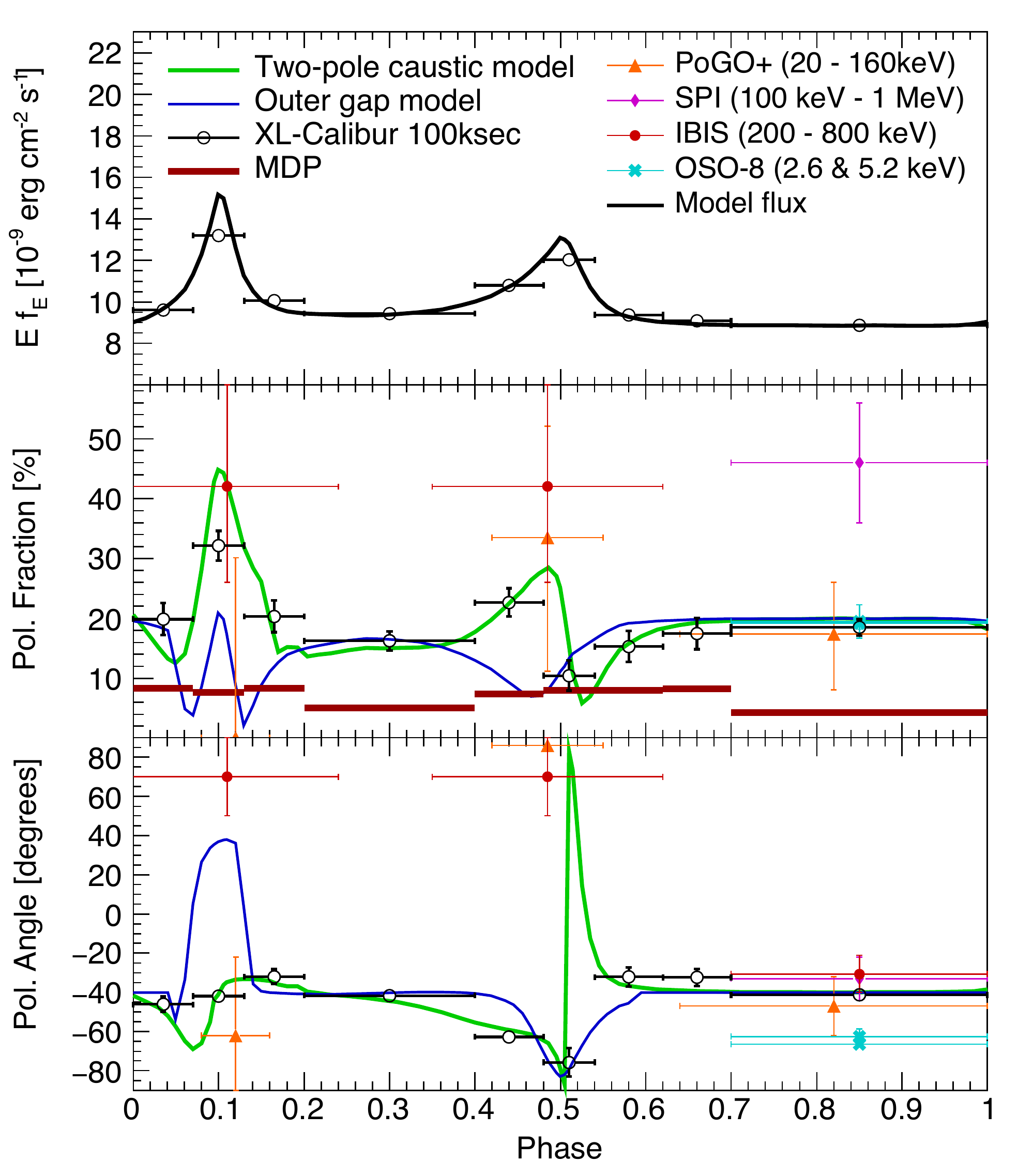}
\end{center}
\caption{\label{f:crab} Simulated phase-resolved {\it XL-Calibur} (15-80 keV) results for a 100~ksec on-source observation of the Crab nebula and pulsar (black data points and dark red 99\% confidence level MDPs) together with models from~\cite{2004ApJ...606.1125D}, and results
from {\it PoGO+}~\cite{2017NatSR...7.7816C,2018MNRAS.477L..45C},
SPI~\cite{2008Sci...321.1183D,2013ApJ...769..137C},
IBIS~\cite{2008ApJ...688L..29F} and OSO-8~\cite{1978ApJ...220L.117W}. {\it XL-Calibur} will deliver definitive results on the hard X-ray polarisation and will allow clear discrimination between models.
An atmospheric column density of 7~g/cm$^2$ is assumed.
}
\end{figure}

{Determining the polarisation properties of the high-energy emission
from isolated pulsars is a powerful tool for investigating their
magnetospheres and in constraining the location of emission sites.} The
high-energy emission is most likely due to charged particles, which are
accelerated within ``gaps'' in the magnetosphere where strong electric
fields can develop. Synchrotron emission arises from e$^{+}$e$^{-}$ pairs
produced in the resulting electromagnetic cascades, {and for the
Crab pulsar appears in the soft and hard X-ray bands~\cite{Harding-2008-ApJ}.
Key questions concern} where these gaps are formed and where the
emission sites are, i.e. in the polar-cap
gap~\cite{1971ApJ...164..529S}, slot gap~\cite{1983ApJ...266..215A},
outer gap~\cite{1986ApJ...300..500C}, variations
there-of~\cite{2004ApJ...606.1125D,2007ApJ...670..677T}, or inside a
current sheet in the equatorial plane~\cite[e.g.][and references
therein]{2018ApJ...857...44K}. For the latter
scenario,~\cite{2016MNRAS.463L..89C} predict an anticorrelation between
flux and polarisation with 15\% (on-pulse) and 30\% (bridge)
polarisations.  {The phase-resolved X-ray polarisation fraction and
angle variations for synchrotron models of the Crab pulsar depend primarily on
whether the emission emanates from inside or outside the light cylinder~\cite{Harding-2017-ApJ}.  Such identification of the emission region
locale is key to understanding the structure and inner workings of the
pulsar magnetosphere.}

A 100~ksec {\it XL-Calibur} on-source observation of the Crab will
measure the phase-resolved polarisation fraction and angle with
exquisite accuracy (Fig.~\ref{f:crab}), enabling model tests with
unprecedented sensitivity. The analysis of the emission from the
spatially extended Crab nebula is complicated, but this caveat does not
apply to the pulsed emission as it originates from a small emission
region comparable to or smaller than the diameter of the light cylinder
($\sim$3000 km), {and is easily isolated from the nebular signal
through timing analyses}.  For a 100~ksec on-source observation of the
Crab nebula and pulsar, 
the 99\% confidence level MDP is 2.3\%. The {\it XL-Calibur} results will
substantially improve on the results of {\it PoGO+}~\cite{2017NatSR...7.7816C,2018MNRAS.477L..45C} and will be complementary
to observations at lower ({\it OSO-8}, {\it IXPE}, and {\it eXTP}) and
higher energies ({\it AstroSat, Hitomi, INTEGRAL}).

%
%
\section{Discussion}
\label{s:discussion}
{\it XL-Calibur} builds on the heritage of the {\it X-Calibur} mission to significantly extend X-ray polarimetry in the 15-80 keV energy band. 
This energy range is an ideal complement to the {\it IXPE} mission (2--8~keV) scheduled for launch in 2021.
{\it XL-Calibur} improves over {\it X-Calibur} by using the {\it FFAST} mirror with a 3--10 times larger
effective area than the {\it InFOC$\mu$S} mirror (resulting in a collection area of $\sim$300 cm$^2$ at 15 keV), and lower background rates resulting from the use of thinner CZT detectors, and improved anticoincidence shielding. The {\it XL-Calibur} approach combines several strengths:
\begin{itemize}
\item {\bf High detection efficiency and low background:} {\it XL-Calibur} detects $\sim$70\% of the scattered photons with a high modulation factor of $\sim$0.5 at all energies, and signal rates exceeding the background rate for $>$200~mCrab sources.
\item {\bf Energy resolution:} \markp{{\it XL-Calibur} achieves an energy resolution of $\sim$3 keV FWHM at 15 keV (dictated by electronic noise) and $\sim$5.9 keV at 40 keV (dictated by the intrinsic CZT energy resolution and Compton scattering energy losses).}
\item {\bf Small systematic errors:} The rotation of the polarimeter during observations allows residual systematic effects, arising from, e.g., variations in the CZT detector response, to be corrected for. 
\end{itemize}

As described in Table~\ref{tab:bgrates}, a reduction in background by a factor of $\sim$6 (25) is predicted for 0.8~mm thick CZT detectors and the {\it X-Calibur} CsI(Na) anticoincidence shield (new compact BGO shield).   
When presenting the science programme, a background level ten times lower than that measured during the 2018/2019 {\it X-Calibur} balloon flight is conservatively assumed. The {\it X-Calibur} flight occurred during maximum background conditions
(close to solar minimum), whereas upcoming flights will occur during
more favourable conditions approaching/around solar maximum. A further reduction in background rates can therefore be expected. 
\begin{figure}[t]
\begin{center}
\includegraphics[width=.99\linewidth]{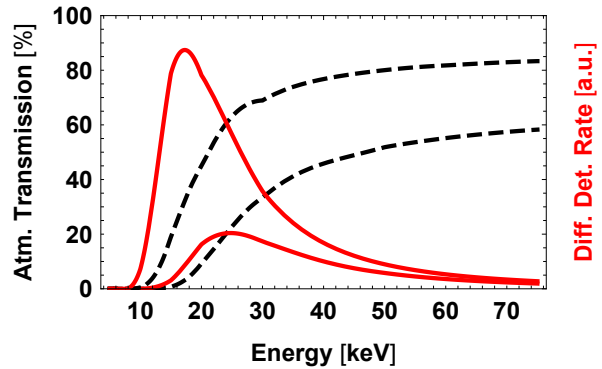}
\end{center}
\caption{\label{f:alt} Atmospheric transmission (dashed black lines) and
photon detection rates for a $dN/dE\propto E^{-3}$ energy spectrum
(solid red lines) for observations at 48.8~km (160~kft) altitude (upper curves, atmospheric column density 1.1 g/cm$^2$) and 39.6~km (130~kft) (lower curves, atmospheric column density 3.2 g/cm$^2$), respectively. 
The development of higher-altitude balloons would lower the energy
threshold of the observations from $\sim$15 keV (39.6~km) to 
$\sim$10~keV (48.8~km). Both graphs assume observations at 70$^{\circ}$ elevation. The XCOM photon cross-sections database provided 
by the National Institute of Standards and Technology~\cite{Berg:2010} has been used.
}
\end{figure}

Competing hard X-ray polarimeter designs use Gas Pixel Detectors (GPDs) similar 
to the ones used on {\it IXPE}, or 
Time Projection Chambers (TPCs) similar to the
ones developed for {\it PRAXyS}~\cite{2016NIMPA.838...89I}.
In both cases, a high detection efficiency can be achieved by optimising the
gas composition and pressure~\cite[e.g.][]{2012ExA....34..463T}.
The trade-offs between the competing techniques involve  
$(i)$ the energy bandpass, 
$(ii)$ the detection efficiency, 
$(iii)$ the energy-dependent effective modulation factor, 
$(iv)$ the energy-dependent background rate, and 
$(v)$ systematic errors. 
{\it XL-Calibur} excels in $(i)$, $(ii)$, $(iii)$, and $(v)$.
On a satellite mission an {\it XL-Calibur}-type polarimeter can cover a broad energy bandpass from $\sim$3 to $>$80~keV. The detection efficiency is near 100\% and the modulation factor is high ($\sim$0.5) over the entire energy range. 
The systematic errors are small and well understood ~\cite{Abarr_2020,Abarr:19b}.
Regarding $(iv)$ both GPD and TPC polarimeters can distinguish photo-electron events from background events by reconstructing track image features.
Referring back to Fig.~\ref{fig:relcont}, hard X-ray GPD and TPC polarimeters are expected to have less (or no) neutron-induced backgrounds, 
but similar or higher gamma-ray induced backgrounds, depending on the shielding.
Considering the photoelectric and Compton cross-sections across the  {\it XL-Calibur} energy band, the {\it XL-Calibur} polarimeter potentially has higher polarisation sensitivity at the upper end of the band, while GPD and TPC polarimeters using appropriate gas mixtures/pressure may have
better sensitivity at the lower end of the band.

Further improvements in {\it XL-Calibur}'s performance could come from the development of higher-altitude balloons. 
The current 1.1 million cubic metre (40 million cubic feet) zero-pressure balloons can carry the 2.1~tonne (4700~lbs)  
{\it XL-Calibur} payload to an altitude of $\sim$40~km (130~kft).
Hard X-ray astronomy would greatly benefit from the development of higher altitude balloons. 
As an example, Fig.~\ref{f:alt} compares the atmospheric transmission at 
$\sim$40~km float altitude to that at 
$\sim$49~km (160~kft) float altitude. 
The higher altitudes would lower the low-energy cutoff from $\sim$15~keV to 
$\sim$10~keV. Assuming a typical energy spectrum $dN/dE\propto E^{-3}$, 
the higher altitude would increase the rate of detected source 
photons by a factor of three. 

An {\it XL-Calibur}-type polarimeter could be used on a stand-alone
satellite borne Small Explorer (SMEX) or Medium Explorer (MIDEX) mission.
The ideal mission would combine the hard X-ray polarimetric capabilities
of an {\it XL-Calibur}-type polarimeter with a REDSOX-type soft 
X-ray polarimeter and intermediate energy polarimeters like those
of {\it IXPE}, {\it eXTP}, or {\it PRAXyS}.
Such a broadband X-ray polarimetry
mission has been proposed~\cite{2019arXiv190710190J,2019arXiv190409313K},
to enable the simultaneous measurement of the 
polarisation of several emission components -- a long-awaited tool for 
precision tests of source emission and geometry models.
%
%
%
\section*{Acknowledgements}
{\it XL-Calibur} is funded by the NASA APRA program under contract number 80NSSC18K0264.
We thank the McDonnell Center for the Space Sciences at Washington University in St. Louis for funding of an early polarimeter prototype, as well as for funds for the development of the ASIC readout. HK acknowledges NASA support under grants 80NSSC18K0264 and NNX16AC42G.
The X-ray measurements of the mirror were performed at BL20B2 at SPring--8 with the approval of the Japan Synchrotron Radiation Research Institute (JASRI) (proposal numbers 2014B1092, 2015A1274, 2016A1035, 2019B1221, and 2020A1298). Support from JSPS KAKENHI (grant numbers 19H01908,
19H05609, and 19K21886) is acknowledged.
KTH authors acknowledge support from the Swedish National Space Agency (grant number 199/18). MP also acknowledges support from the Swedish Research Council (grant number 2016-04929). Dana Braun (Washington University in St. Louis) is thanked for producing Fig.~\ref{f:czt}.

%
\bibliographystyle{elsarticle-num}

%
%
\end{document}